# Quantum mechanics without quanta

arXiv:1507.02113 [quant-ph]

**Sergey A. Rashkovskiy**

*Institute for Problems in Mechanics of the Russian Academy of Sciences, Vernadskogo Ave., 101/1, Moscow, 119526, Russia*

*Tomsk State University, 36 Lenina Avenue, Tomsk, 634050, Russia*

*E-mail: rash@ipmnet.ru, Tel. +7 906 0318854*

**Abstract**    In this paper, I argue that light is a continuous classical electromagnetic wave, while the observed so-called quantum nature of the interaction of light with matter is connected to the discrete (atomic) structure of matter and to the specific nature of the light-atom interaction. From this point of view, the Born rule for light is derived, and the double-slit experiment is analysed in detail. I show that the double-slit experiment can be explained without using the concept of a "photon", solely on the basis of classical electrodynamics. I show that within this framework, the Heisenberg uncertainty principle for a "photon" has a simple physical meaning not related to the fundamental limitations in accuracy of the simultaneous measurement of position and momentum or time and energy. I argue also that we can avoid the paradoxes connected with the wave-particle duality of the electron if we consider some classical wave field - "an electron wave" - instead of electrons as the particles and consider the wave equations (Dirac, Klein-Gordon, Pauli and Schrödinger) as the field equations similar to Maxwell equations for the electromagnetic field. It is shown that such an electron field must have an electric charge, an intrinsic angular momentum and an intrinsic magnetic moment continuously distributed in the space. It is shown that from this perspective, the double-slit experiment for "electrons", the Born rule, the Heisenberg uncertainty principle and the Compton effect all have a simple explanation within classical field theory. The proposed perspective allows consideration of quantum mechanics not as a theory of particles but as a classical field theory similar to Maxwell electrodynamics.

**Acknowledgments**
Funding was provided by Tomsk State University competitiveness improvement program.

**Preface**

Reading this document, it is necessary to keep in mind that each its part was prepared and published (or submitted to a Journal) as a separate paper. However, arXiv moderator considered that these papers cannot be submitted to arXiv as a separate documents and should be incorporated into a replacement of arXiv:1507.02113. Which was done. For this reason, each part of the document has its own References and its own numbering the Sections, equations and figures.

**PART 1: THE NATURE OF THE WAVE-PARTICLE DUALITY OF LIGHT**

**1   Introduction**

There are two opposing perspectives on the nature of light.

According to the first perspective, i.e., orthodox, light has both wave- and particle-like properties: in some experiments (diffraction and interference experiments), light behaves as a classical electromagnetic wave, while in other experiments (interaction with matter), light behaves as a flux of particles, i.e., photons. This property of light was called the wave-particle duality. Directly, the wave-particle duality of light can be observed in many optical experiments; the duality is manifested in the observation that light interacting with matter induces discrete events (clicks of detector or the appearance of spots on a photographic plate), which are interpreted as the interactions of single photons; however, after prolonged exposure, these discrete events merge into a single continuous pattern, well described by classical electrodynamics or even classical optics [1,2].

Such a perspective on the nature of light is now considered to be the official one, being described in all textbooks and monographs on quantum physics and optics. From this point of view, the wave-particle duality is considered to be a fundamental property of nature. At the same time, the wave-particle duality is one of the greatest mysteries of modern physics because it is impossible to imagine a physical object that has the properties of both a wave (object continuously distributed in a space) and a particle (object localised in a small region of space). Most clear and impressively, the paradoxes associated with the wave-particle duality of light are manifested in Young's double-slit experiment and in Wiener's experiments with standing waves: the closing of one of the slits in the double-slit experiment changes the "behaviour" of the photons that pass through the open slit, while the installation or removal of a mirror in the Wiener experiments results in a "spatial redistribution" of photons in the incident beam [3]. These paradoxes gave rise to a whole new direction in quantum physics - the interpretation of quantum mechanics [3-12].

Experiments indicate that the wave properties of light cannot be explained as a result of the interaction of photons in the beam because they are manifested, even for a very weak light source, when "photons fly alone" [1,2]. This behaviour indicates that each individual photon should have wave properties. As a result, the representation that the photon interferes with itself has appeared.

The discovery of the wave-particle duality of light raised numerous fundamental questions among physicists, the most difficult of which are the following: (i) What is a photon? (ii) How may one and the same matter (light) have such incompatible properties in terms of classical physics: both wave and particle? These questions were considered by A. Einstein, who made several attempts to answer them [13-15]. On 12 December 1951, Albert Einstein wrote to M. Besso: "*All these 50 years of conscious brooding have brought me no nearer to the answer to the question: What are light quanta?*"

Despite the fact that the concept of a "photon" has become generally accepted, following A. Einstein, we can say that even now, after more than 100 years since the introduction of the concept of the wave-particle duality in physics, we are unable to answer the fundamental question: what is a photon?

Quantum electrodynamics, which is based on the Copenhagen interpretation that postulates the wave-particle duality, approaches this problem formally, providing a mathematical formalism for the calculation of the various phenomena at the atomic and subatomic levels and, at the same time, abandoning the classical images of waves and particles. In fact, currently, we have an elegant mathematical theory that does not represent the object it actually describes.

Compatibility between the wave and corpuscular properties of light in quantum theory is achieved by using a probabilistic interpretation of optical phenomena: the probability $p$ of finding a photon at some point in space is proportional to the intensity of the classical light wave $I \sim \mathbf{E}^2$ at this point, calculated on the basis of the methods of wave optic [16]:

$$p \sim \mathbf{E}^2 \qquad (1)$$

where $\mathbf{E}$ is the strength of the electric field of the classical light wave.

Equation (1) is a mathematical formulation of the wave-particle duality because the probability $p$ refers to the particle (photon), while the intensity $I \sim \mathbf{E}^2$ refers to the wave, which, in many cases, can be calculated by the methods of classical wave optics.

The rule (1), underlying the photonic interpretation of experiments with light, is an independent postulate of quantum mechanics.

We will call equation (1) for brevity "the Born rule for light" by analogy with the same type of Born rule describing the probability of finding a nonrelativistic quantum particle.

Thus, the description of optical experiments within the framework of the photon hypothesis consists of two independent parts: (i) the Maxwell equations describing the propagation of light and (ii) the Born rule (1). Precisely, the Born rule (1) results in a paradox in explaining the wave-particle duality of light [3].

At the same time, there is another point of view, according to which there are no photons; in this perspective, light is a continuous classical electromagnetic wave that is completely described by Maxwell equations, while the so-called "quantum properties" of light are "manifested" only during the interaction of light with matter, which is connected to the specific nature of this interaction. This point of view is now shared by only a small part of physical community, but at a different time, this point of view was supported by many authoritative physicists, such as M. Planck, E. Schrödinger, W.E. Lamb, Jr., A. Lande, E.T. Jaynes et al. For example, Planck was a proponent of a semi-classical theory, in which only the atoms and their interactions were quantised, while the free fields remained classical. We note that even N. Bohr, one of the authors of the Copenhagen interpretation of quantum mechanics and the most implacable defender of the Copenhagen interpretation, at the beginning, with his usual enthusiasm, was also an opponent of light quanta [17], although he implicitly placed photons into the basis of his mechanistic model of the hydrogen atom. The most consistent opponents of the concept of photons were the Nobel Prize Winner Willis E. Lamb, Jr. [18,19] and Alfred Lande [20]. E.T. Jaynes had questioned the need for a quantum theory of radiation [21].

Indeed, many quantum phenomena, traditionally described by quantum electrodynamics, can in fact be explained within the framework of so-called semiclassical theory, in which atoms are described by the wave equations of quantum mechanics (Schrödinger, Dirac, etc.), while light is described by classical electrodynamics without quantisation of the radiation. These phenomena include the photoelectric effect [22,23], the Compton effect [24-27], the Lamb shift [28-31], radiative effects [29-32], spontaneous emission [28,31,32], the Hanbury Brown and Twiss effect [41,42], etc. In recent years, this point of view is again attracting the attention of physicists [21,33-38] who are not satisfied with existing interpretations of quantum mechanics.

I completely share this position: many well-known experimental facts, which are traditionally interpreted as a manifestation of the wave-particle duality of light, can in fact be explained on the basis of purely wave representations of light without involving the "photon" concept. In this article, I will describe the arguments in support of this point of view.

In the present paper, the author begins a series of articles involving the justification and development of the point of view that the photon does not exist and that light is a classical wave field described by the Maxwell equations. In doing so, the Maxwell equations and the wave equations (Schrödinger, Dirac, etc.) describing the atoms are found to be sufficient for a complete and consistent description of the majority of well-known experiments involving light and matter.

## 2  Wave-particle duality of light

2.1 Interaction of a classical electromagnetic wave with an atom

The "corpuscular" properties of light are manifested in its interaction with matter. Precisely an interpretation of the results of experiments on the light-matter interaction has led to the emergence of the concept of a "photon" as a "particle of light". In the early years of the development of quantum theory, it was believed that only the quantisation of the light field can explain the processes of light-matter interaction; however, later, it became clear that these phenomena can be described within the limits of the so-called semiclassical theories [22-38,43-47].

Note that many of these results have been known since the dawn of quantum mechanics; however, in view of the rapid and certainly remarkable successes of quantum electrodynamics, they have not received proper attention and development.

Common to all the semi-classical theories is that light is considered as a classical electromagnetic wave, which is described by Maxwell's equations, while the atoms of matter, with which this wave interacts, are described by the Schrodinger equation or other wave equations (Klein-Gordon, Pauli Dirac), depending on the degree of detail of the described process.

The question of why the atoms of matter are described by the wave equations will be discussed in subsequent articles of this series.

If the interaction of light (classical electromagnetic waves) with an atom is weak compared to the intra-atomic interactions (which is typical for the majority of experiments with light), then the process can be described within the limits of perturbation theory [39], according to which the probability of excitation (ionisation) of an atom in the field of a monochromatic electromagnetic wave for time $\Delta t$ is equal to

$$w\Delta t = bI\Delta t \qquad (2)$$

where $w$ is the probability of excitation (ionisation) of atoms per unit time; $I \sim \mathbf{E}^2$ is the intensity of the classical electromagnetic wave at the location of the atom; and $b$ is a constant depending on the characteristics of the atom, the frequency of the incident electromagnetic wave and the states between which the transition of the atom occurs under the action of the electromagnetic wave.

Expression (2) is sufficient to explain, without using the concept of a "photon", the many processes of light-matter interaction, in which the so-called "quantum" properties of light are manifested.

Note that unlike the Born rule (1), expression (2) is not a postulate; rather, it follows from the exact theory (exact in the sense in which, for example, the Schrodinger equation is an exact equation describing the atom).

Hereinafter, I will show that based on expression (2) and taking into account the discrete (atomic) structure of matter, it is easy to explain the "wave-particle duality" of light, remaining within the realm of classical electrodynamics, while the concept of a "photon" becomes superfluous.

2.2 Born rule for light

Suppose there is a wave field $\mathbf{E}$, which represents, generally, the result of the interference of classical electromagnetic waves. If we place a detector into this field, then an excitation of an arbitrarily chosen atom of the detector occurs with probability (2); this interaction would be perceived as a triggering of the detector. Thus, the random excitation of the atom (the detector operation) would occur under the action of a continuous electromagnetic wave and is not related to the discrete (quantum) structure of the light. Taking into account the fact that the detector consists of a plurality of atoms, the sequence of discrete events, i.e., excitations of different atoms of the detector, would be observed under the action of the electromagnetic wave, which can occur either in the form of clicks of the detector or in the form of the appearance of spots on a photographic plate. These events occur randomly in space and time, but after a prolonged exposure, they form a continuous, non-random, deterministic macroscopic picture.

Let us introduce the probability $P_-(t)$ that a randomly selected atom located in the light field is not excited (ionised) for time $t$. Obviously, $P_-(t)$ satisfies the equation

$$\frac{dP_-}{dt} = -wP_- \tag{3}$$

with the initial condition

$$P_-(t_0) = 1 \tag{4}$$

where $t_0$ is the time of the beginning of irradiation of the atom by the electromagnetic wave; $w$ is determined by expression (2).

The solution of Eqs. (3) and (4) is as follows

$$P_-(t) = \exp\left(-\int_{t_0}^{t} w(t')dt'\right) \tag{5}$$

Accordingly, the probability that the atom would be excited (ionised) at time $t$ is

$$P_+(t) = 1 - P_-(t) \tag{6}$$

When the electromagnetic wave is weak and satisfies the condition

$$\int_{t_0}^{t} w(t')dt' \ll 1 \tag{7}$$

one approximately obtains

$$P_+(t) \approx \int_{t_0}^{t} w(t')dt' \tag{8}$$

Taking into account expressions (2) and (8), for an atom in a weak electromagnetic wave, one obtains

$$P_+(t) \sim \mathbf{E}^2 \tag{9}$$

If we use the "photonic" interpretation of the observed process, each excitation of the atom would be interpreted as a "photon" hitting the point that is the location of the atom. In this case, we need to assume that a "photon" hitting the given point in space inevitably causes a click of the detector or the appearance of a spot on the photographic plate. In this case, we can discuss the probability of finding the "photon" at the given point in space. Such an analysis implies that we apply a discreteness of matter (detector), consisting of atoms, to the "structure" of light: instead of discrete matter, we consider it a continuous one, while instead of a continuous electromagnetic (light) wave, we consider the flux of discrete quanta – "photons". As a result, we interpret a single event (a click of the detector or the appearance of a spot on the photographic plate) as hitting the "photon" at some point in space filled with continuous matter. In this case, the number of excited atoms of matter (the detector) is considered to be equal to the number of "photons" that arrived at the detector. Let the total number of atoms in a selected volume $V$ of matter (e.g., on a selected surface of the photographic plate) be equal to $N$ and be uniformly distributed thereon. Let us choose a small volume $dV$, in which $dN = \frac{N}{V}dV$ atoms are located. Then, for time $t$ within volume $dV$, $dN_+ = P_+ dN$ atoms would be excited:

$$dN_+ = P_+ \frac{N}{V} dV \tag{10}$$

Because we interpret the excitation of the atoms as photons hitting the atoms, the number of photons entering the volume $dV$ would be determined by expression (10). Accordingly, the probability of a photon entering into a given volume $dV$ is

$$pdV = \frac{P_+}{\int P_+ dV} dV \tag{11}$$

where the integral is taken over the entire volume (e.g., the entire surface of the photographic plate).

Thus, the probability density to detect a photon at a given point in space is given by

$$p = \frac{P_+}{\int P_+ dV} \sim P_+ \tag{12}$$

Using expressions (5) and (6), one obtains

$$p(t) \sim 1 - \exp\left(-\int_{t_0}^{t} w(t')dt'\right) \tag{13}$$

In the case of weak electromagnetic waves satisfying condition (7), we have

$$p \sim \int_{t_0}^{t} w(t')dt' \tag{14}$$

Taking into account expression (2), one obtains

$$p \sim \mathbf{E}^2 \tag{15}$$

This expression is none other than the Born rule for "photons" (1).

The above analysis leads to the following conclusions: (i) the Born rule for light (1) is a trivial consequence of quantum mechanics if the electromagnetic wave is considered as a classical field, while the matter (detector) is considered as consisting of discrete atoms; (ii) the Born rule (1) is valid only for weak electromagnetic waves and a relatively short exposure time, satisfying condition (7). For strong electromagnetic waves or for a longer exposure time for which condition (7) is not satisfied, the Born rule should be replaced by the stricter rule (13), according to which the dependence of probability $p$ on $\mathbf{E}^2$ is more complicated:

$$p(t) \sim 1 - \exp\left(-\gamma \mathbf{E}^2 \Delta t\right) \tag{16}$$

where, taking into account (2), the following is introduced

$$\int_{t}^{t+\Delta t} w(t')dt' = \gamma \mathbf{E}^2 \Delta t \tag{17}$$

and $\gamma$ does not depend on $\mathbf{E}^2$.

Thus, instead of the postulated Born rule (1), the more general rule (2), following from quantum mechanics and classical electrodynamics, should be used; the more general rule does not lead to paradoxes, such as the wave-particle duality. If we accept the rule (2) as a basis, then the Born rule (1) will immediately follow from it, as a reasonable approximation, which is valid, however, only for weak waves or short exposure times.

We see that the direct application of the Born rule (1) as the primary principle leads to the wave-particle duality, while the more general rule (2) does not lead to paradoxes in interpreting the experimental data.

## 2.3 Double-slit experiment

Double-slit experiments [1,2], along with experiments involving standing waves [48-51], are direct and evident "demonstrations" of the wave-particle duality of light and the related paradox. Numerous attempts to explain these experiments by unusual (non-classical) motion of point "photons" were unsuccessful [4-12].

Here, I show that the double-slit experiment can be easily explained in terms of classical electrodynamics if we take into account the discrete structure of matter (a screen, a detector) and the specific nature of the interaction of light with matter, which is described by the Schrodinger equation (or other wave equation of quantum mechanics).

Let a weak classical electromagnetic wave pass through the diffractive device, such as a diffraction grating or a system of slits in an opaque screen, and impinge on the surface of the photoactive substance (conditionally, a photographic plate), the atoms of which can be excited by light.

For simplicity, I assume that the photographic plate represents a single layer of atoms, in which the atoms are placed randomly with surface density $n$ (number of atoms per unit surface). The atoms are the quantum objects, which are described by the Schrodinger equation, whereas the light is a classical electromagnetic wave. The light intensity can be calculated using classical optics and is considered to be known on the surface of the photographic plate.

We are interested in the interaction of a weak light wave with matter, when the "photons" impinge on the surface of the photographic plates one by one. It is these conditions realized in the double-slit experiments [1,2] that demonstrate "the wave properties of individual photons".

The probability of excitation (ionisation) of any atom on the surface of photographic plates is described by expression (2).

Generally speaking, relation (2) itself explains the "wave-particle duality of light"; nevertheless, for a clear "demonstration" of the "wave-particle duality of light", we will perform the direct calculation of the interaction of light with a photographic plate, assuming that the excitation (ionisation) of an atom is perceived as blackening of the appropriate point of the photographic plate.

The calculation proceeds using the Monte Carlo method: at each time, the probability of excitation of each unexcited (non-ionised) atom will be calculated. The calculation for each atom is continued until its excitation (ionisation) occurs.

The rate of atomic excitation $w$ does not depend on the concentration of atoms and is determined only by the intensity of the radiation at a given point of the screen. If the radiation intensity does not change with time, then the law of excitation of atoms will be similar to that of radioactive decay. In particular, the probability of excitation at time $t$ of an atom located at a point on the screen with a given value $w$ is

$$P_+(t) = 1 - \exp(-wt) \tag{18}$$

Taking into account (2), one obtains

$$P_+(t) = 1 - \exp(-bIt) \tag{19}$$

Let us introduce the nondimensional time

$$\tau = bI_0 t \tag{20}$$

where $I_0$ is the maximum intensity of light on the screen (photographic plate).

In this case, the probability of excitation of the atom for time $t$ at the points on the screen having the intensity of light $I$ is

$$P_+(t) = 1 - \exp\left(-(I/I_0)\tau\right) \tag{21}$$

As an example, I consider the calculation of the double-slit experiment. In this case, the distribution of the light intensity on the screen is given by the well-known expression [51]

$$I(z) = I_0 \cos^2\left(\frac{d}{b}x\right)\left(\frac{\sin x}{x}\right)^2 \tag{22}$$

where

$$x = \frac{\pi b}{\lambda H} z, \quad z = H \sin\theta$$

$b$ is the width of the slits; $d$ is the distance between the slits; $\lambda$ is the wavelength of light; $H$ is the distance from the slits to the screen; and $\theta$ is the angular coordinate.

For the calculation of the double-slit experiment using the Monte Carlo method, I create a system of randomly and uniformly distributed points $i = 1,...,N$ in a given area $L_z \times L_y$, simulating the screen. These points are considered as the atoms of the material of the screen. I use the average distance between atoms as a length scale; in these units, the concentration of the atoms on the surface of the screen is equal to unity.

At each moment of time $\tau$ for each yet unexcited atom, the probability $P_{+i}$ is calculated by expressions (21) and (22); at the same time, the random number $R_i \in [0,1]$, $i = 1,...,N$, is generated by a random number generator. If the condition $R_i \leq P_{+i}$ is satisfied, then the given atom is considered to be excited, and it is depicted by a black dot. Unexcited atoms are not depicted.

The results of the calculations of the process of the "accumulation of photons" on the screen for different moments of time $\tau$ at $\frac{\pi b}{\lambda H} = 0.03$ and $d/b = 5$ are shown in Fig. 1 (left). The markers on the graph on the right are the histogram obtained by treating the corresponding picture on the left; the line on the graph on the right is the theoretical dependence (22), predicted by classical optics.

Comparing Fig. 1 with the real picture of the "accumulation of photons" on the photographic plate in the double-slit experiment [2], we see that the calculated pattern is completely consistent with the experimental interference pattern and the model completely reproduces the results of these experiments: at short exposure times or at low intensities of the light, the random system of dots on the screen appears, which can be interpreted as the places of the "fall of the photons" on the screen, although no photons were considered in our model. With increasing exposure time or intensity of light, these dots form clear interference patterns corresponding to the theoretical distribution of intensity, as follows from wave optics. Closing one of the slits, we obtain, in accordance with the considered calculation scheme, exactly the picture predicted by classical optics. In other words, we do not find the "wave-particle duality" of light and the related paradox.

Thus, we see that the double-slit experiments can be easily interpreted in terms of wave optics and quantum mechanics, based on the Schrödinger equation.

Obviously, this method also allows for reproduction of the results of the Wiener experiments with standing waves for low light intensity and, in particular, Lippmann fringes.

Note that according to expression (21), the change in light intensity $I_0$ does not result in changes in the pattern in Fig. 1; only the time scale is changed: at high light intensity, the same pattern is reached at a shorter exposure time. At high light intensity or at a long exposure time, condition (7) may be violated, and the pattern on the photographic plate will be different from the predictions of wave optics. This means that the simple Born rule (1) for light in these cases ceases to work, and we need to use the common rule (21).

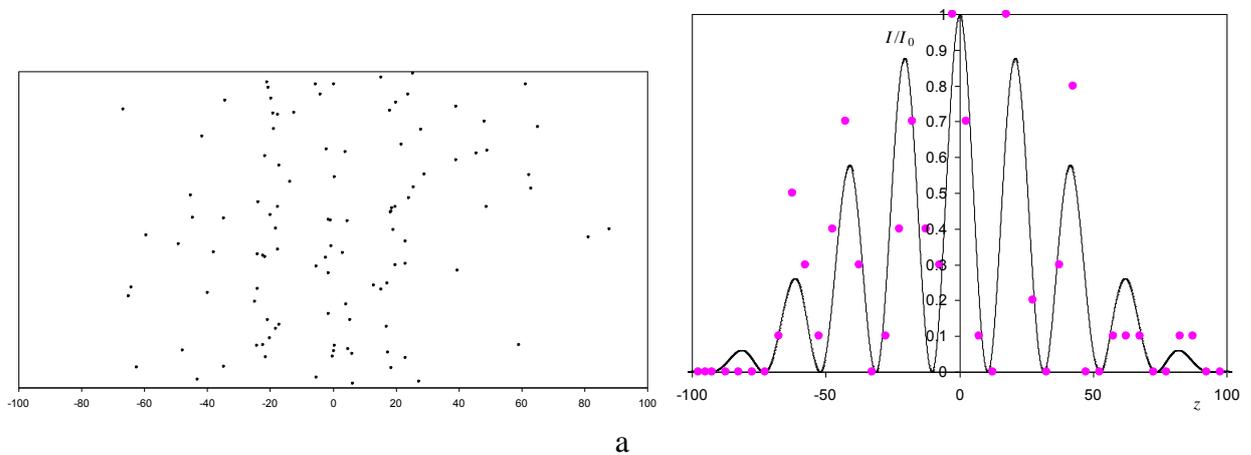

a

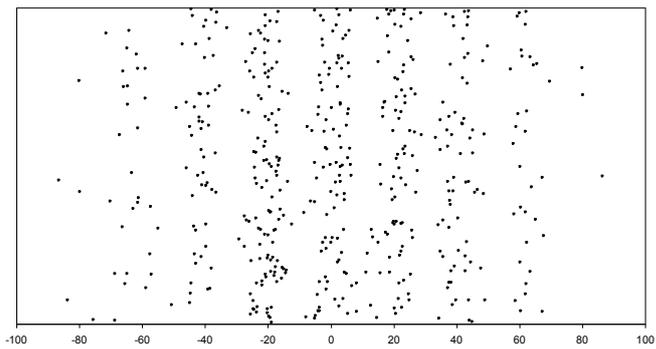 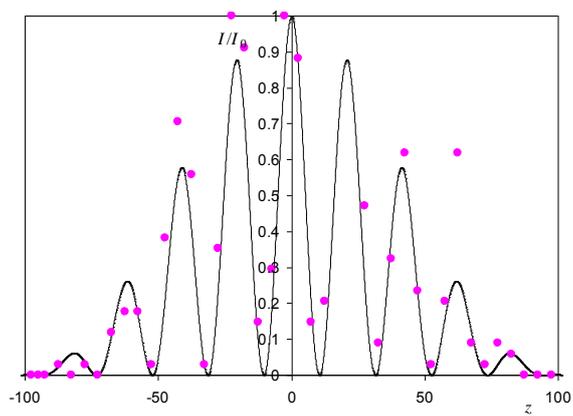

b

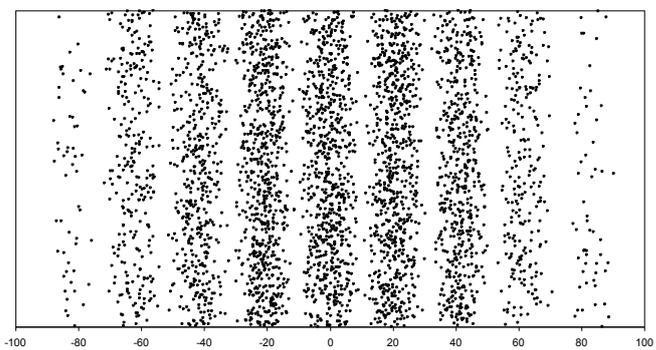 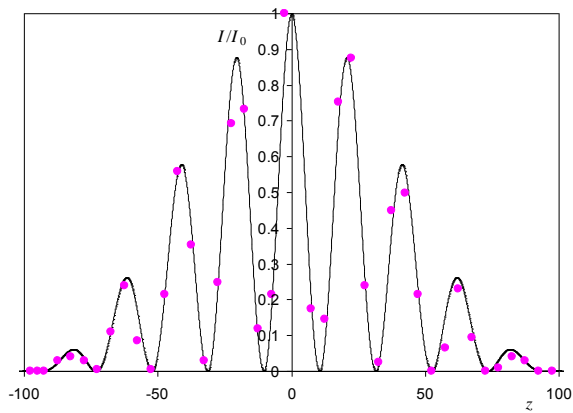

c

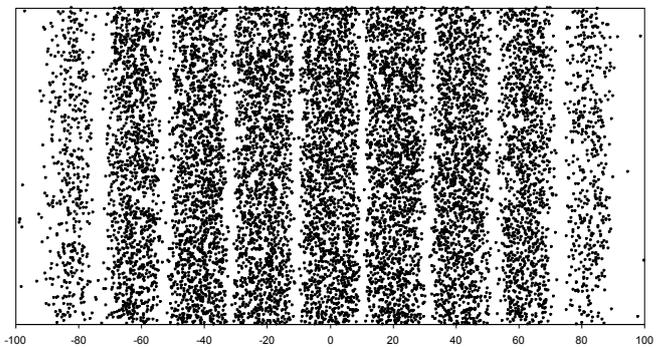 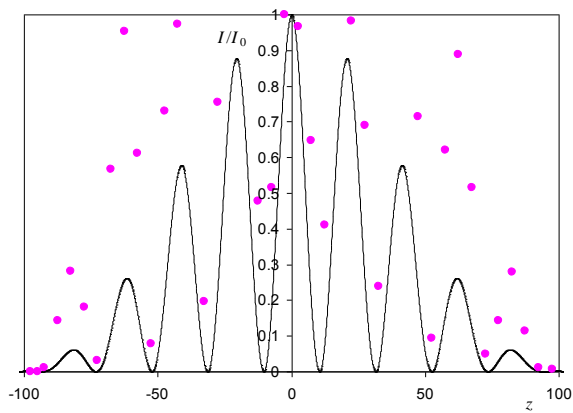

d

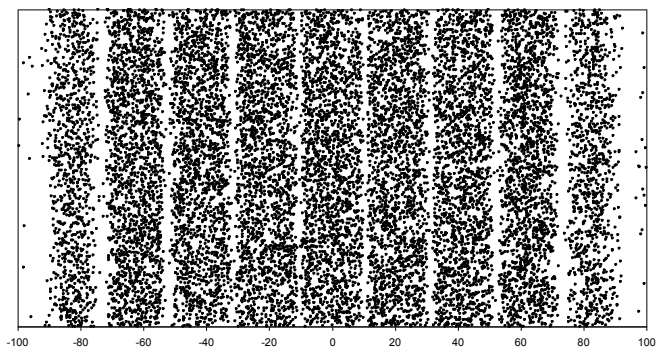 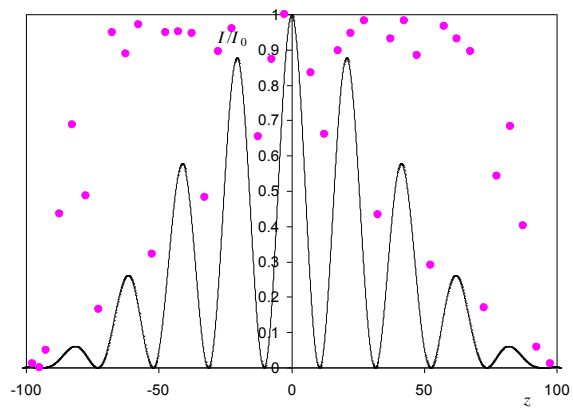

e

**Fig. 1** (Colour online) Interference pattern buildup (left) and the corresponding distribution functions of dots on the screen (right) for different exposure times $\tau$, obtained using the Monte Carlo simulation of the interaction of light with the detecting screen. a) $\tau = 0.02$ (100 "photons"); b) $\tau = 0.1$ (424 "photons"); c) $\tau = 1$ (3452 "photons"); d) $\tau = 10$ (11600 "photons"); e) $\tau = 30$ (14530 "photons"). The markers on the graphs (right) show the histograms obtained by treating the corresponding picture on the left picture; the line is the theoretical dependence (22), predicted by classical optics.

For an arbitrary exposure time, the probability "to detect a photon" at a given point of the photographic plate is determined by expression (12), which, taking into account expression (21), can be written as

$$\frac{p}{p(0)} = \frac{1 - \exp(-(I/I_0)\tau)}{1 - \exp(-\tau)} \quad (23)$$

where $p(0)$ is the probability "to detect a photon" in the middle of the photographic plate (at $I = I_0$). The ratio $p/p(0)$ in the double-slit experiments with the "single photons" plays the same role as the ratio $I/I_0$ in the optical experiments. In particular, at the short exposure times $\tau \ll 1$, we obtain from expression (23) the Born rule: $p/p(0) \approx I/I_0$.

At a long-term exposure, the diffraction pattern will qualitatively appear as that predicted by wave optics, i.e., a system of periodic fringes, but quantitatively, it will be significantly different from both the predictions of wave optics (the line in Fig. 1, right) and the predictions based on the simple Born rule (1). The quantitative agreement with the wave theory will be observed only at relatively short exposure times, when condition (7) is satisfied. Note that the quantitative disagreement with the wave theory is observed also at very short exposure times (Fig. 1a and 1b); however, if for a long exposure time, this difference is due to the approximate nature of the Born rule (1), for short exposure times, this difference is connected to the random scatter due to the small number of recorded events. If one performs a large number of similar tests with a short exposure time and averages the results of these tests, then for $\tau \ll 1$, the obtained pattern will exactly correspond to the predictions of classical optics (22) and the Born rule for light (1). This result is confirmed by Fig. 2 (left), which presents the results of the calculations for $\tau = 0.1$ averaged over 10 statistical realisations. A similar result corresponding to the long exposure ($\tau = 5$) is shown in Fig. 2 (right). The lines in Fig. 2 (right) show expression (23) corresponding to $\tau = 0$ (it coincides with expression (22)) and $\tau = 5$. In this case, the distribution of dots on the screen is found to be significantly different from that predicted by classical optics (22) and the Born rule (1); however, the distribution is well described by the general expression (23). For comparison, Fig. 3 shows the normalised distribution function calculated by expression (23) for different exposure times $\tau$.

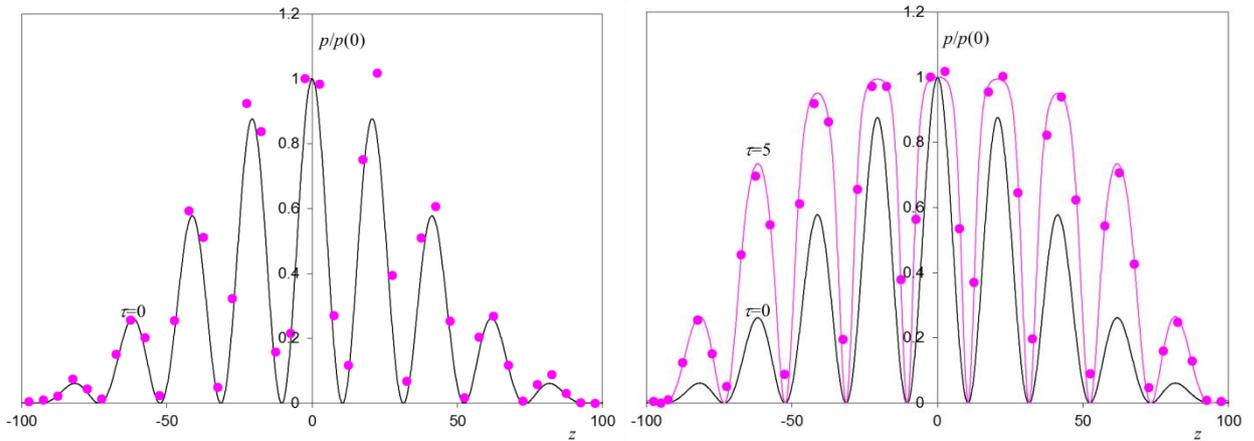

**Fig. 2** (Colour online) Comparison of the results of Monte Carlo simulations using expression (21) for $\tau = 0.1$ (left) and $\tau = 5$ (right) and averaged over 10 statistical realisations (markers), with expression (22) (predicted by classical optics (line $\tau = 0$)) and with dependence (23) (predicted by classical electrodynamics, taking into account expression (2) (line $\tau = 5$)). The data shown in Fig. 1 correspond to one particular test.

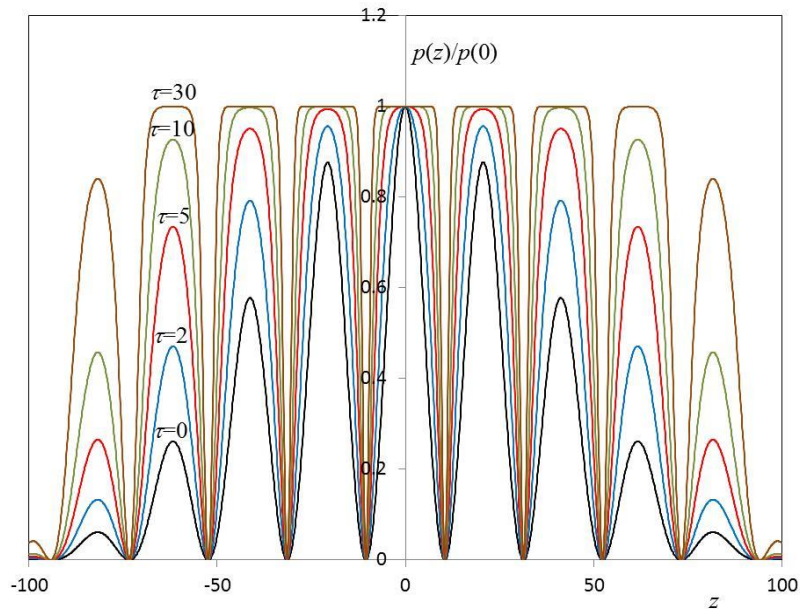

**Fig. 3** (Colour online) Normalised distribution functions of the number of "photons" calculated using expression (23) for different exposure times $\tau$.

Note that the theoretical dependence (22) (line $\tau = 0$ in Fig. 1-3) describes the distribution of dots on the screen, which must be obtained within the "photon" theory if we assume that the probability of finding a photon at a given point is determined by the Born rule (1) and all the photons falling on the photographic plate cause its blackening with the same probability. Thus, according to the "photon" theory, the normalised distribution of dots on the screen should not depend on the exposure time (at least as long as the fraction of excited atoms on the surface of the screen is small). In contrast, according to the "photon-free" theory, when light is considered as a classical electromagnetic wave, the distribution of dots on the screen (23) will substantially depend on the exposure time, and already at $\tau > 1.5$, the distinction from the "photon" theory (22) will be clearly visible (Figs. 2 and 3). This fact can be used for experimental verification of the theory.

## 3 Concluding remarks

The following are the most important "quantum" effects for which the theoretical calculation does not require the quantisation of radiation:

1. Double-slit experiment and the Wiener experiments with standing waves.
2. Compton effect. In quantum mechanics, the Compton effect occupies a special place. This effect is considered to be direct proof of the existence of photons. Exactly after the discovery of the Compton effect and its explanation based on photonic representations, many physicists began to perceive photons as the real physical objects. In the canonical, for quantum mechanics, explanation of the Compton effect, it is considered an elastic scattering of photons by free electrons. In this case, photons are considered as massless relativistic particles that have energy $\hbar\omega$ and momentum $\hbar\omega/c$. Such an explanation of the Compton effect is included in all textbooks and books on physics.

However, even at the dawn of quantum mechanics, the Compton effect was explained without using the concept of the photon [24-27,53,54]: light was considered as a classical electromagnetic wave, while electrons were described by Klein-Gordon or Dirac equations. From the considered points of view, the approach [26] based entirely on classical electrodynamics is of particular interest. In that paper, the classic electric current, which creates the scattered electromagnetic wave, was calculated on the basis of the solution of the Klein-Gordon equation. At the same time, in papers [24,25], the wave equation was used only as a tool for calculating the matrix elements; this approach makes the obtained results more formal and less clear from the physical point of view. A pictorial explanation of the Compton effect was proposed by E. Schrödinger [53] on the basis of purely wave representations: he considered light as a classical electromagnetic wave and drew an analogy between the scattering of this wave on the de Broglie wave and the Bragg scattering of light on ultrasonic waves considered by L. Brillouin.

3. Photoelectric effect. All the known regularities of the photoelectric effect are explained if light is considered as a classical electromagnetic wave, whereas atoms are described by the wave equation, e.g., Schrödinger or Dirac [22,23,54]. From perturbation theory, it follows [39,40] that a threshold effect occurs at the transition of an atomic electron into the continuum: if the frequency of the incident light is less than the threshold frequency $U_i/\hbar$, where $U_i$ is the ionisation potential of the atom, then ionisation does not occur and $w = 0$. Otherwise, the ionisation probability is determined by expression (2). This completely explains the photoelectric effect without invoking the concept of a "photon".

4. Hanbury Brown and Twiss effect [41,42]. This effect has a simple classical explanation [43-47] if expression (2) is used and we assume that the components of the electric field vector of the light wave **E** are random variables and have a Gaussian distribution. In this case, each click of the detector is considered to be the result of the excitation of one of the atoms under the action of light.

5. Interaction of an intense laser field with an atom. One of the striking examples of how "photons" appear in theory that does not consider the quantisation of light is the Keldysh theory [55], which describes the multiphoton and tunnel ionisation of atoms in an intense laser field. In the Keldysh theory, atoms are described by the Schrodinger equation, while light is considered as a continuous classical electromagnetic wave. Nevertheless, the obtained continuous solution of the Schrodinger equation is interpreted from the standpoint of "photonic" theory, which allows for the conclusion that, under certain conditions, "multiphoton" ionisation of an atom occurs when the atom "simultaneously absorbs several photons". Such an interpretation is based on the fact that the solution of the Schrodinger equation contains components with a phase factor $n\hbar\omega$, where $n = 1,2,...$, which is interpreted as a result of the simultaneous absorption of $n$ "photons". From the mathematical point of view, these "multiphoton" terms are simply the conventional terms of the expansion of solutions in the Fourier series, and in any other ("non-quantum") theories, they would be perceived as quite trivial. It is paradoxical that radiation is not quantised in the Keldysh theory and in its solution and is considered as a continuous wave; the "photons" appear only at the stage of interpreting the results of the solution. This situation is typical for many processes in which the "quantum effects" are manifested.

6. The Heisenberg uncertainty principle, along with the Born rule and the complementarity principle, is the basis of the Copenhagen interpretation of quantum mechanics. The Heisenberg uncertainty principle is considered as a quantitative justification of the wave-particle duality. This justification is achieved by interpreting the uncertainty relations as the constraints imposed by nature on the precision of simultaneous measurements of the position and momentum of a quantum object.

Let us consider the classical electromagnetic wave packet $\mathbf{E}(\mathbf{r},t)$. From the properties of the Fourier transform, which is the basis of an elementary proof of the uncertainty relations (see, e.g., [16,51]), it follows that

$$\Delta x \Delta k_x \geq \frac{1}{2} \qquad (24)$$

where $\Delta x$ is the characteristic width of the wave packet; $\Delta k_x$ is the characteristic width of the range of wave numbers $k_x$ of monochromatic waves entering into the packet.

Obviously, the uncertainty relation (24) does not have the mystical meaning that is ascribed to the Heisenberg uncertainty principle in the Copenhagen interpretation. The relation simply states a mathematical fact: the greater the width of the wave packet, the smaller the range of wave numbers of monochromatic waves entering into the wave packet and vice versa. Multiplying expression (24) by the Planck constant and using the formal expression $\mathbf{p} = \hbar \mathbf{k}$, we obtain the Heisenberg uncertainty relation

$$\Delta x \Delta p_x \geq \frac{1}{2}\hbar \qquad (25)$$

This relation can be interpreted as a formal connection between the spatial width of the wave packet and a range of momentum of the "photons" entering into the packet. If the energy of the classical wave packet $\mathbf{E}(\mathbf{r},t)$ is less than the energy of one "photon" $\hbar\omega$, then it is necessary to resort to the probabilistic interpretation (1), and expression (25) should be interpreted as the limitations on the accuracy of simultaneous measurements of the position and momentum of the "photon". As we have shown above, the explanation of many "quantum effects" does not require the use of the concept of a "photon" as a real physical object. Therefore, in reality, expression (25) contains no more meaning than expression (24). A similar conclusion can be made regarding the Heisenberg uncertainty relation for time and energy.

We see that many so-called quantum phenomena (even the most iconic for quantum theory) can be described in detail without the quantisation of radiation within the images that are contained already in classical field theory.

From this point of view, the "photons" are the result of the incorrect interpretation of optical phenomena, where, instead of the actual process of the interaction of a continuous classical electromagnetic wave with a detector, which has a discrete (atomic) structure, a fictitious system is considered, in which the flux of discrete particles of light (the "photons") interacts with a continuous (structureless) detector. In this case, the real discrete structure of the detector is replaced by a fictitious discrete structure of radiation (e.g., light).

We see that the use of the concept of "photons" as the real "particles of light" in all the above cases is not required to explain the interaction of electromagnetic waves with a detector. However, we can save the notion a "photon" as a synonym for discrete events – clicks of a detector, appearance of spots on a photographic plate, etc. In other words, the photon should not be seen as a physical object ("particle of light"); rather, the photon is only a discrete event that is observed in experiments, which actually arises from the interaction of a continuous electromagnetic wave with discrete objects (atoms of a detector).

Here, it is possible to reproach the author that he rejects the notion that photons are real physical objects while considering atoms as quantum objects, thereby involuntarily considering that atoms emit and absorb electromagnetic energy in the form of discrete portions (quanta) due to jump-like transitions. We note only that the wave-particle duality of nonrelativistic matter will be considered in the next papers, in which this contradiction will be eliminated. We will develop this concept for other forms of matter and show that the emission of light by atoms also occurs continuously, while the observed discrete spectrum of atom emission has a simple classical explanation. We will also consider other "quantum" effects and show that they can also be explained without the quantisation of radiation.

## PART 2: THE NATURE OF THE ELECTRON

### 1   Introduction

Initially, electrons were discovered as cathode rays (J. Plücker, 1859) and similar to light, they were interpreted as precisely as rays.

Subsequently, it became clear that cathode rays are deflected in the electric and magnetic fields (J.B. Perrin, 1885) and therefore, they carry an electrical charge. According to the mechanistic representations of the time, the electric charge has been associated with the particles and thus, cathode rays began to be considered as a flux of negatively charged electric particles, which were called electrons (J.J. Thomson, 1897). This view became predominant in physics; subsequently, it became the only view that was recognized.  The discovery of the Zeeman effect (P. Zeeman, 1896) and its explanation by H.A. Lorentz based on electronic representations largely contributed to this view. Additional confidence that the electron is a particle that always has the same electric charge was obtained from experiments on determining the electron's charge, first conducted by R.A. Millikan (1910-1912), that clearly indicated that the electric charge is quantized and can take only values that are multiples of the elementary electric charge $e$ (modern value $e$=1.6021766208(98)×10$^{-19}$ C). In experiments on the deflection of cathode rays in electric and magnetic fields (J.J. Thomson, 1897), the results of which have been interpreted from the corpuscular perspective, the $e/m_e$ ratio was determined; that ratio has enabled the determination of the mass of the particle-electron $m_e$ (modern value $m_e$ =9.10938356(11)×10$^{-19}$ kg). J.A. Lorenz's theory of the electron (1892-1909), which describes the behaviour of electrons in electric and magnetic fields as the movement of electrically charged point particles, was the peak of this classical period.

Thus, the electron was considered as a material point with the mass $m_e$ and an electric charge $-e$ that obeys the laws of classical mechanics (including the theory of relativity) and is described by the Maxwell-Lorentz electrodynamics.

Experiments by J.J. Thomson indicated that electrons are included in the composition of ordinary matter.

Based on this perspective and on Lorentz's theory of the electron, it was possible to explain a large number of observed phenomena, including many electrical and optical properties of various media, plasma phenomena, etc. Thus, nobody doubted the electron's existence as a negatively charged elementary particle.

Simultaneously, the classical theory of the electron was immediately confronted by a number of contradictions that it was unable to overcome, including the following: (i) the infinite electromagnetic energy of the point electron; (ii) the infinite density of the mass and electric charge of a point particle with a finite mass and a finite electric charge; (iii) the "self-acceleration" of a point's electric charge caused by the inverse action of its own radiation field (a point's electric charge after passing through any field would be infinitely "self-accelerated"; the absurdity of this result is obvious).

After E. Rutherford suggested the planetary model of the atom (E. Rutherford, 1911) and N. Bohr built a mechanistic model of the hydrogen atom based on the planetary model (N. Bohr, 1913), the number of contradictions in the theory of the electron increased, adding exotic (from the perspective of classical physics) properties such as the existence of a discrete set of allowed orbits of the electron in an atom and jump-like transitions from one allowed orbit to another with the emission or absorption of an energy quantum (i.e., a photon). Bohr's postulates contradicted all known laws of classical physics; in particular, the postulate that an accelerating electron in a stationary orbit does not radiate electromagnetic waves was inexplicable in terms of classical electrodynamics. Despite these contradictions, Bohr's theory worked and gave the correct prediction for the hydrogen atom; classical mechanics and electrodynamics had been unable to do the same.

At least formally, the contradictions of the Rutherford-Bohr theory and classical physics (mechanics and electrodynamics) were successfully overcome only as the result of a radical revision of the concepts of particles and waves. Louis de Broglie advanced his revolutionary hypothesis (L. de Broglie, 1924): each particle-electron (later, this hypothesis was extended to all kinds of particles) corresponds to a certain wave (i.e., the de Broglie wave); the length of wave $\lambda$ is associated with particle momentum $p$ by a ratio similar to that obtained from Einstein's work for light quanta (photons):

$$\lambda = \frac{h}{p} \qquad (1)$$

where $h$ is the Plank constant.

From a formal perspective, the de Broglie hypothesis was a simple generalization on the non-relativistic forms of matter of the wave-particle duality of light, introduced by Einstein to explain the photoelectric effect (A. Einstein, 1905). However, the de Broglie hypothesis actually represented a return to the idea of cathode rays (which are now called the de Broglie wave), albeit on a new basis: now, these rays were considered both waves and flux of particles, even if they consisted of a single particle-electron.

The wave properties of electrons were soon discovered experimentally in the scattering of the electron beam on the crystal lattice of the single crystal (C.J. Davisson and L.H. Germer, 1927; G.P. Thomson, 1927): the observed interference pattern in these experiments left no doubt that the flux of electrons has wave properties similar to those of electromagnetic radiation. However, electrons' wave properties could not be attributed to some interaction between the electrons in the beam, because after long-term exposure, even a very weak electron beam in which theoretically the electrons should move one by one creates the same interference pattern. In connection with this, the representation appeared that each electron has the wave properties and can interfere with itself. A similar property is attributed to the photon as a "light particle".

The wave-particle duality of electrons has been demonstrated clearly in the double-slit experiment with single electrons [1]: after passing through the "electron biprism", a weak flux of electrons left discrete bright spots on the detecting screen that "left no doubt" that the beam consists of individual particles (electrons), but after prolonged exposure, the number of these spots become so large that they merge into a practically continuous interference pattern similar to that observed in the double-slit experiment in weak light [2,3], thus "leaving no doubt" that this beam is a wave.

Thus, because nobody refused to view the electron as a point particle, a new fundamental problem was added to the existing problems of the classical theory of the electron: the wave-particle duality.

Soon after the appearance of the de Broglie hypothesis, E. Schrödinger suggested the wave equation (E.R.J.A. Schrödinger, 1926) that described wave field $\psi(\mathbf{r},t)$, which was continuously distributed in space. The spectrum of the eigenvalues of this equation coincided with the optical spectrum of the hydrogen atom that was experimentally observed and predicted by the Bohr theory. It followed that wave field $\psi(\mathbf{r},t)$, which was described by the Schrödinger equation, has some relation to the electron in a hydrogen atom.

Immediately, there was a problem with interpreting wave field $\psi(\mathbf{r},t)$. The first interpretation of wave function $\psi$ was suggested by Schrödinger himself. That interpretation considered $|\psi|^2$ as a measure of the distribution of the electric charge of electron in space. Schrödinger considered it possible to abandon both quantum jumps and corpuscular representations and to consider the electron as a wave packet described by the wave equation.

However, under the influence of N. Bohr's unquestionable authority and his theory of the hydrogen atom—which was considered a reference, claiming that (i) the electron is a point particle and moves around the nucleus and (ii) the electron makes the jump-like transitions from one stationary level to another that are accompanied with an emission or absorption of an energy quantum $\hbar\omega$— Schrödinger's proposed interpretation was rejected. Opponents of Schrödinger's interpretation additionally argued that it would inevitably conflict with Planck's theory of thermal radiation, which requires a discreteness of energy.

A probabilistic interpretation of the wave function was proposed by M. Born (1926). That interpretation solved the electron scattering problem using the Schrödinger equation. Born considered the electron as a point particle, the probability of finding of which in space in the neighbourhood of point $\mathbf{r}$ is proportional to $|\psi|^2$:

$$p(\mathbf{r},t) \sim |\psi(\mathbf{r},t)|^2 \qquad (2)$$

where $p$ is the probability density of finding of the electron at a given point in space.

As mentioned by Born himself, the probabilistic interpretation of the wave function arose under the influence of Einstein's interpretation of light intensity as a measure of the distribution density of photons in space.

The Born rule (2) is *an independent postulate of quantum mechanics* and represents a mathematical formulation of the electron's wave-particle duality because probability $p$ characterizes the particle, whereas field $\psi(\mathbf{r},t)$ characterizes the wave.

Attempts were made to derive the Born rule from within standard quantum mechanics. For example, W.H. Zurek presented a derivation of the Born rule based on a mechanism termed "environment-assisted invariance" [4-6]. However, it should be noted that this derivation cannot be considered convincing. More likely, it is an attempt to find a class of systems for which the Born rule is obeyed.

Thus, one can say that quantum mechanics is composed of two independent parts: (i) the wave equation (for example, the Schrödinger equation) describing a certain field $\psi(\mathbf{r},t)$; and (ii) the Born rule (2), which enables interpretation of the solutions to the wave equation and a comparison with the experimental results.

Although the interpretation proposed by M. Born better corresponded to Bohr's representations and to postulates of his theory of the hydrogen atom, it required a philosophical justification.

The Heisenberg uncertainty principle (W.K. Heisenberg, 1927) was the first such "justification"; it imposed restrictions on the accuracy of the simultaneous measurement of coordinate $\mathbf{r}$ and momentum $\mathbf{p}$ of a quantum particle (electron), which has considered as a material point:

$$\Delta x \Delta p_x \geq \frac{1}{2}\hbar \qquad (3)$$

Similar relations hold for coordinates $y$ and $z$. Subsequently, a similar uncertainty relation was formulated with respect to time $t$ and energy $E$:

$$\Delta t \Delta E \geq \frac{1}{2}\hbar \qquad (4)$$

The uncertainty principle is considered as the fundamental restriction imposed by nature on the accuracy of the measurement of a point particle's (electron's) physical parameters and this accuracy cannot be improved in any manner.

The complementarity principle suggested by N. Bohr (1927) became the second, already philosophical, "justification" of the wave-particle duality according to which the wave and the corpuscular description of the micro-objects are additional to each other. Thus, the particle's (electron's) wave and corpuscular properties are never shown simultaneously: in some experiments, we can observe only the wave properties, whereas in others, we can observe only the corpuscular properties. Moreover, it is claimed that the quantum object cannot be presented using the classical images of a wave and a particle because they reflect our representations of a macrocosm. As a result, the so-called Copenhagen interpretation of quantum mechanics (1927) appeared based on (i) the Born probabilistic interpretation of the wave function, (ii) the Heisenberg uncertainty principle and (iii) the Bohr complementarity principle.

Despite the fact that the Copenhagen interpretation is considered as "official" and accepted by the majority of the physics community, it creates dissatisfaction in connection with its refusal of the usual classical images of a wave and a particle and the loss of the physical theory's obviousness.

This dissatisfaction resulted in the appearance of alternative interpretations of quantum mechanics [7-16], some of which appear fantastic. Without dwelling on those interpretations, we will note that for various reasons, none of them seriously pretend to provide a final, consistent physical interpretation of quantum mechanics. Here, we will mention only one of those interpretations: the instrumentalist interpretation of quantum mechanics, which is often equated with eschewing all interpretation. It is summarized by the sentence "Shut up and calculate!", which is attributed to various physicists (P. Dirac, R. Feynman, D. Mermin).

Previously, when discussing the wave equation, we kept in mind the Schrödinger equation. It should be noted that the Schrödinger equation was not derived from first principles: indeed, it was guessed. Subsequently, operator formalism was developed, which enables one to obtain the Schrödinger equation for any mechanical system if the Hamilton function is known for its classical analogue. However, this again was not a derivation of the Schrödinger equation from first principles; it served simply as method of transition from the description of the relevant classical system to the description of its quantum analogue. This method is excused only because it practically always leads to the correct result. Using this method, a relativistic generalization of the Schrödinger equation—the Klein-Gordon equation—was obtained (O. Klein, W. Gordon, V.A. Fok, 1926).

One more of the electron's riddles arose after the discovery of a new property: spin (G.E. Uhlenbeck and S. A. Goudsmit, 1925). From the analysis of the spectroscopic data, it followed that the electron-particle behaves like a charged "spinning top" with its own angular momentum, equal to $(1/2)\hbar$, and its own magnetic moment, equal to the Bohr magneton $\mu_B = \dfrac{e\hbar}{2cm_e}$. The problem was that it makes no sense to talk about an own-rotation-of-point particle: if we assign to the electron the radius $r_e = \dfrac{e^2}{m_e c^2}$ ("classical radius of electron"), then the spherical electron, having own angular momentum $(1/2)\hbar$, would have a velocity on its own surface that is greater than the speed of light, which contradicts the theory of relativity. A second problem associated with electron spin was that the spin gyromagnetic ratio of the electron was 2 times higher than the gyromagnetic ratio of the system of charged particles when the magnetic moment is associated with the motion of electric charges. A third problem associated with spin was that to explain the experimental data, it was necessary to accept the condition (W.E. Pauli, 1924, 1927) that the projection of an electron's intrinsic angular momentum (spin) on any chosen axis can take only two values $\pm\frac{1}{2}\hbar$, which is impossible for the ordinary vector.

Attempts to find a mechanistic explanation for the spin and associated with it the magnetic moment of the electron proved unsuccessful and ultimately it was agreed to consider the spin as an intrinsic, purely quantum property of the electron, not reducible to the mechanical motion of electric charges that are attributed to the electron, similar to the rest mass and the electric charge. For an explanation of numerous quantum effects (primarily, the anomalous Zeeman effect), Pauli introduced the exclusion principle (W.E. Pauli, 1925), according to which one quantum state cannot be occupied by more than one electron simultaneously. Because the Pauli exclusion principle is associated with the spin of the particle, this problem (a paradox in terms of classical physics) can also be relegated to the range of problems arising from the presence of the electron spin $\frac{1}{2}\hbar$. Because there are no restrictions similar to the Pauli exclusion principle in classical systems, this principle is considered to be a purely quantum property of the electron that cannot be explained based on classical representations.

To describe the electron in an external magnetic field while taking spin into account, Pauli proposed an equation (W.E. Pauli, 1927) that was a generalization of the Schrödinger equation and that took into account that the spin's transformation properties are similar to those already known by the time of the transformational properties of the electron's orbital angular momentum. Pauli simply postulated the value of the intrinsic magnetic moment of the electron entered into this equation, having taken it from the Uhlenbeck-Goudsmit theory.

The final stage in the construction of a quantum theory of the electron was P. Dirac's discovery of the relativistic wave equation for a particle with a ½ spin (P.A.M. Dirac, 1928), which in the non-relativistic approximation automatically turns into the Pauli equation with the correct gyromagnetic ratio. This was one of the triumphs of Dirac's electron theory.

It is now believed that the Dirac equation provides the most complete description of the electron. From the brief description above, we can see that in the process of developing the modern theory of the electron, contradictions and problems gradually accumulated and have survived to the present day.

Thus, we can state that despite the fact that the concept of the electron as a particle has existed in physics for more than 150 years, there are still no answers to the questions of what is an electron, what does it look like, what are its properties connected to, etc. However, this does not prevent us from using quantum mechanics and calculating various physical processes involving electrons with high (and sometimes fantastic) accuracy. Thus, as with the photon, although we now have a well-composed theory of the electron, we do not imagine the object that this theory describes.

In my opinion, the key problem is the wave-particle duality of the electron, having solved which we can hope to get answers to other questions.

The double-slit experiment with the "single electrons" [1] and similar experiments seemingly convince us that an electron has both wave and corpuscular properties.

Similar experiments with light have also "convinced" us that light has both wave and particle properties. However, in [17-19], it is shown that many processes of the interaction of light and matter, including the double-slit experiment, can easily be explained in terms of a pure wave theory of light, if one considers the discrete (atomic) structure of matter (detector) and the specific nature of the interaction of light waves with atoms, which is described, for example, by the Schrödinger equation. From this perspective, physical objects such as photons are completely superfluous and the quantization of the radiation is considered only as a mathematical technique used to solve numerous problems. Using this approach, in [17-19] it is shown that the "Born rule for light" is a trivial consequence of the Schrödinger equation, occurring only for relatively short exposure times, whereas for long-term exposure it is necessary to use a more general nonlinear rule. The Heisenberg uncertainty principle as applied to light from these positions is not a fundamental limitation of the accuracy of measuring the parameters of a particle-photon, as claimed by the Copenhagen interpretation; instead, it only fixes a well-known (from classical optics) property of the wave packet: the smaller the width of the wave packet, the wider the range of wave numbers of monochromatic waves entering into that packet [17,18].

Comparing similar experiments with electrons and light (in the first place, the double-slit experiments [1-3,20]), we arrive at the conclusion that experiments with "single electrons" can also be explained if the electron beam is considered not as a flux of particles but as some classic (continuous) wave, which we conventionally call the electron wave. In that situation, the discrete events observed in the experiments (clicks of the detector or the appearance of the points on the detecting screen) are solely the result of the interaction of the continuous electron wave with the discrete atoms that comprise the detector.

In this paper and in the following papers of this series, I will show that many properties of atoms and "electrons" have a natural and intuitive explanation in the framework of classical physics, if we abandon the electron as a particle; instead, however, we will consider the electron wave as a classical continuous wave field described by the Dirac equation.

In this paper, the main properties of the electron wave will be established and a classical explanation of the numerous phenomena considered iconic in the mythology of quantum mechanics will be given.

## 2 Quantum mechanics as a classical field theory

### 2.1 Maxwell-Klein-Gordon field

I propose to consider the following perspective: *there are no electrons as particles, but instead there is an electron wave, which is a real classical wave field*, in the sense that the wave is continuous in space and time. From this perspective, the Dirac equation is the equation of the electron field, similar to Maxwell's equations for the classical electromagnetic field. The Dirac equation gives the most complete description of the behaviour of electron waves under any conditions, whereas other wave equations (Klein-Gordon, Pauli and Schrödinger), which are also electron-field equations, give only an approximate description of this field corresponding to various limiting cases. We assume that the wave function (scalar, spinor or bispinor, depending on the detalization of the process) described by these equations is a real physical parameter like the strengths of electric and magnetic fields in classical electrodynamics. In my opinion, we should not be frightened by the complex values of the wave function because the same wave equations can be written as a system of real equations for real field functions. Eventually, it is possible to use the hydrodynamic forms of writing wave equations that operate within real parameters (see below).

From this perspective, it makes sense to operate without the parameters that characterize the particle, for example, the mass of the electron $m_e$, but instead to use the parameters that characterize the wave, for example, the natural frequency of the electron wave $\omega_e$, which is considered the fundamental physical constant: $\omega_e = 7.763440716 \times 10^{20}$ rad/s. The mass of the electron $m_e$ appearing in the theory of electron and its applications, along with the natural frequency of the electron wave $\omega_e$, are connected by the relation

$$m_e = \hbar \omega_e / c^2 \tag{5}$$

Thus, we consider frequency $\omega_e$ as a primary parameter, whereas the mass of electron $m_e$ is an auxiliary parameter that is used for convenience in some (primarily classical) applications.

Describing the electron wave as a real physical field, we first shall neglect properties such as spin, which will be considered later. In this section, we consider the electron wave described by the Klein-Gordon equation, which has the necessary relativistic invariance. For the same electron wave, a transition from the "Klein-Gordon description", which uses wave function $\Psi$, to the "Schrödinger description", which uses wave function $\psi$, occurs with the help of the following ratio:

$$\Psi = \exp(-i\omega_e t)\psi \qquad (6)$$

Therefore, considering the electron field in the Klein-Gordon approximation, we will simultaneously keep in mind the Schrödinger approximation.

The most general formulation of classical field theory can be given on the basis of the variational principle [21,22]:

$$\delta \int \mathscr{L} \, d\Omega = 0 \qquad (7)$$

where $\mathscr{L}$ is the Lagrangian density; $d\Omega = cdt\,dx\,dy\,dz$.

For the Maxwell-Klein-Gordon field, the Lagrangian density has the form [22]

$$\mathscr{L} = \frac{\hbar c^2}{2\omega_e}\left(g^{\mu\nu}(\partial_\mu \Psi^* + i\alpha_0 A_\mu \Psi^*)(\partial_\nu \Psi - i\alpha_0 A_\nu \Psi) - \frac{\omega_e^2}{c^2}\Psi^*\Psi\right) - \frac{1}{16\pi} F_{\mu\nu} F^{\mu\nu} \qquad (8)$$

where as usual, $A_\mu$ is the 4-potential of the electromagnetic field;

$$F_{\mu\nu} = \partial_\mu A_\nu - \partial_\nu A_\mu \qquad (9)$$

is the tensor of the electromagnetic field, the components of which are the strengths of an electrical $\mathbf{E}$ and a magnetic $\mathbf{H}$ field: $F_{\mu\nu} = (\mathbf{E},\mathbf{H})$ [21]; $g_{\mu\nu} = (1,-1,-1,-1)$ is the metric tensor of planar space-time; $\partial_\mu = \frac{\partial}{\partial x^\mu}$, $\mu = 0,1,2,3$; $x^\mu = (ct,\mathbf{r})$, $\alpha_0 = \frac{e}{\hbar c} > 0$ is the constant of interaction—the fundamental constant that characterizes the interaction of two classical fields, i.e., the Klein-Gordon (electron) field and the electromagnetic field.

In three-dimensional form, the Lagrangian density (8) takes the form

$$\mathscr{L} = \frac{\hbar c^2}{2\omega_e}\left(\frac{1}{c^2}\left(\frac{\partial \Psi^*}{\partial t} + i\alpha_0 c\varphi\Psi^*\right)\left(\frac{\partial \Psi}{\partial t} - i\alpha_0 c\varphi\Psi\right) - \right.$$
$$\left. -(\nabla\Psi^* - i\alpha_0 \mathbf{A}\Psi^*)(\nabla\Psi + i\alpha_0 \mathbf{A}\Psi) - \frac{\omega_e^2}{c^2}\Psi^*\Psi\right) + \frac{1}{8\pi}(\mathbf{E}^2 - \mathbf{H}^2) \qquad (10)$$

In variational principle (7), the field functions $A_\mu$, $\Psi^*$ and $\Psi$ are considered to be independent.

Taking the variations of Eqs. (8) and (10) with respect to $\Psi^*$, one obtains the Klein-Gordon equation [22]

$$g^{\mu\nu}(\partial_\mu - i\alpha_0 A_\mu)(\partial_\nu - i\alpha_0 A_\nu)\Psi + \frac{\omega_e^2}{c^2}\Psi = 0 \qquad (11)$$

whereas taking the variations with respect to $A_\mu$, one obtains the first couple of the Maxwell equations

$$\partial_\nu F^{\mu\nu} = -\frac{4\pi}{c} j^\mu \qquad (12)$$

where

$$j_\mu = -i\frac{ec^2}{2\omega_e}\left(\Psi^*(\partial_\mu\Psi - i\alpha_0 A_\mu\Psi) - \Psi(\partial_\mu\Psi^* + i\alpha_0 A_\mu\Psi^*)\right) \qquad (13)$$

Variations of Eqs. (8) and (10) with respect to $\Psi$ lead to the complex-conjugate Klein-Gordon equation and therefore are not something new compared to (11).

The second couple of Maxwell equations is connected to the determination of the tensor of the electromagnetic field (9) and takes the form [21]

$$\partial_\sigma F_{\mu\nu} + \partial_\mu F_{\nu\sigma} + \partial_\nu F_{\sigma\mu} = 0 \qquad (14)$$

According to the classical interpretation of Maxwell equation (12), $j^\mu$ is the 4-vector of the electric current densities, connected with the distribution and motion of the electric charges producing the electromagnetic field: $j^\mu = (c\rho, \mathbf{j})$ where $\rho$ is the density of the electric charge, $\mathbf{j}$ is the three-dimensional electric current density.

By definition, (9), the tensor of the electromagnetic field, is antisymmetric: $F_{\mu\nu} = -F_{\nu\mu}$; thus, it follows from Maxwell equation (12)

$$\partial_\mu j^\mu = 0 \qquad (15),$$

which is the continuity equation and describes the law of conservation of the electric charges.

It follows that the classical electron wave field described by wave function $\Psi$ has an electric charge that is continuously distributed in space with density

$$\rho = -i\frac{e}{2\omega_e}\left(\Psi^*\frac{\partial\Psi}{\partial t} - \Psi\frac{\partial\Psi^*}{\partial t} - i2\alpha_0 c\varphi\Psi\Psi^*\right) \qquad (16)$$

and an electric current density

$$\mathbf{j} = i\frac{ec^2}{2\omega_e}\left(\Psi^*\nabla\Psi - \Psi\nabla\Psi^* + i2\alpha_0 \mathbf{A}\Psi\Psi^*\right) \qquad (17)$$

Thus, *the first basic distinction of the electron field from the electromagnetic field is that the electron field has an electric charge and electric current that are continuously distributed in space, cannot be reduced to the motion of any charged particles, and are intrinsic properties inherent to the electron field.*

The physical meaning of wave function $\Psi$ follows from expressions (16) and (17): it is a measure of the distribution of the electric charge and electric current of an electron wave in space that practically coincides with Schrödinger's initial interpretation.

The classical Maxwell-Klein-Gordon field is the nonlinear, self-consistent field, which is described by a set of field functions ($A_\mu, \Psi$) satisfying equations (11)-(14).

Similar to any classic field, the Maxwell-Klein-Gordon field has energy and momentum continuously distributed in space, which are described by the symmetric energy-momentum tensor [22]

$$T_{\mu\nu} = \frac{\hbar c^2}{2\omega_e} \Big( (\partial_\mu \Psi^* + i\alpha_0 A_\mu \Psi^*)(\partial_\nu \Psi - i\alpha_0 A_\nu \Psi) + (\partial_\nu \Psi^* + i\alpha_0 A_\nu \Psi^*)(\partial_\mu \Psi - i\alpha_0 A_\mu \Psi) -$$
$$- g_{\mu\nu} \left[ g^{\sigma\gamma}(\partial_\sigma \Psi^* + i\alpha_0 A_\sigma \Psi^*)(\partial_\gamma \Psi - i\alpha_0 A_\gamma \Psi) - \frac{\omega_e^2}{c^2}\Psi^*\Psi \right] \Big) + \qquad (18)$$
$$+ \frac{1}{4\pi}\left( -g^{\sigma\gamma} F_{\mu\sigma} F_{\nu\gamma} + \frac{1}{4} g_{\mu\nu} F_{\sigma\gamma} F^{\sigma\gamma} \right)$$

The term

$$T_{\mu\nu}^{em} = \frac{1}{4\pi}\left( -g^{\sigma\gamma} F_{\mu\sigma} F_{\nu\gamma} + \frac{1}{4} g_{\mu\nu} F_{\sigma\gamma} F^{\sigma\gamma} \right) \qquad (19)$$

represents the Minkowski tensor, a well-known energy-momentum tensor in the classical electromagnetic field [21]. The remaining terms of tensor (18) describe the spatial distribution of the energy and momentum of the electron field, taking into account its interaction with the electromagnetic field.

Energy-momentum tensor satisfies the continuity equation

$$\partial_\nu T_{\mu\nu} = 0 \qquad (20),$$

which is the differential form of the conservation laws of energy and momentum for the Maxwell-Klein-Gordon field [21].

As is known [21], component $W = T^{00}$ is the energy density of the field, whereas the components $P^s = \frac{1}{c} T^{s0}$ form the vector of the momentum density of the field $\mathbf{P}$, where $s = 1, 2, 3$.

Using expression (18), for the Maxwell-Klein-Gordon field one can write

$$W = \frac{\hbar c^2}{2\omega_e} \left( \frac{1}{c^2}\left( \frac{\partial \Psi^*}{\partial t} + i\alpha_0 c\varphi\Psi^* \right)\left( \frac{\partial \Psi}{\partial t} - i\alpha_0 c\varphi\Psi \right) + \right.$$
$$\left. + (\nabla\Psi^* - i\alpha_0\mathbf{A}\Psi^*)(\nabla\Psi + i\alpha_0\mathbf{A}\Psi) + \frac{\omega_e^2}{c^2}\Psi^*\Psi \right) + \frac{1}{8\pi}(\mathbf{E}^2 + \mathbf{H}^2) \quad (21)$$

$$\mathbf{P} = -\frac{\hbar c}{2\omega_e}\left( \frac{1}{c}\left( \frac{\partial \Psi}{\partial t} - i\alpha_0 c\varphi\Psi \right)(\nabla\Psi^* - i\alpha_0\mathbf{A}\Psi^*) + \frac{1}{c}\left( \frac{\partial \Psi^*}{\partial t} + i\alpha_0 c\varphi\Psi^* \right)(\nabla\Psi + i\alpha_0\mathbf{A}\Psi) \right) +$$
$$+ \frac{1}{4\pi c}[\mathbf{EH}] \quad (22)$$

The term

$$W^{em} = \frac{1}{8\pi}(\mathbf{E}^2 + \mathbf{H}^2)$$

represents a well-known expression for the energy density of the classical electromagnetic field [21], whereas the term

$$\mathbf{S} = \frac{c}{4\pi}[\mathbf{EH}]$$

is the Poynting vector, which describes the energy flux density of the electromagnetic field [21].

2.2 The De Broglie wave

For purposes of illustration, let us consider the solution to the Klein-Gordon equation corresponding to a planar electron wave in free space ($A_\mu = 0$). In this case, we neglect not only an external electromagnetic field but also its own electromagnetic field, which creates an electron wave because of its distributed electric charge.

The solution takes the form

$$\Psi = u\exp(-i(\omega t - \mathbf{kr})) \quad (23),$$

where $u$ is the constant amplitude of the wave; $\omega$ and $\mathbf{k}$ are the frequency and the wave vector of the electron wave.

Substituting expression (23) in the Klein-Gordon equation (11) at $A_\mu = 0$, one obtains

$$\omega^2 = \omega_e^2 + c^2\mathbf{k}^2 \quad (24)$$

For long electron waves satisfying the condition

$$\mathbf{k}^2 \ll \frac{\omega_e^2}{c^2} \quad (25),$$

one obtains approximately

$$\omega = \omega_e + \frac{1}{2}\frac{c^2}{\omega_e}\mathbf{k}^2 \qquad (26)$$

It is in this limit case that the Klein-Gordon equation turns into the Schrödinger equation, which describes the slowly varying amplitude $\psi$ of the electron field (6) in the long-wave approximation.

In cases in which the electromagnetic field can be neglected, expressions (16), (17), (20) and (21) take the following form:

$$\rho = -i\frac{e}{2\omega_e}\left(\Psi^*\frac{\partial \Psi}{\partial t} - \Psi\frac{\partial \Psi^*}{\partial t}\right) \qquad (27)$$

$$\mathbf{j} = i\frac{ec^2}{2\omega_e}\left(\Psi^*\nabla\Psi - \Psi\nabla\Psi^*\right) \qquad (28)$$

$$W = \frac{\hbar c^2}{2\omega_e}\left(\frac{1}{c^2}\frac{\partial \Psi^*}{\partial t}\frac{\partial \Psi}{\partial t} + \nabla\Psi^*\nabla\Psi + \frac{\omega_e^2}{c^2}\Psi^*\Psi\right) \qquad (29)$$

$$\mathbf{P} = -\frac{\hbar}{2\omega_e}\left(\frac{\partial \Psi}{\partial t}\nabla\Psi^* + \frac{\partial \Psi^*}{\partial t}\nabla\Psi\right) \qquad (30)$$

For a planar electron wave (23), one obtains

$$\rho = -e\frac{\omega}{\omega_e}|u|^2 \qquad (31)$$

$$\mathbf{j} = -\frac{ec^2}{\omega_e}\mathbf{k}|u|^2 \qquad (32)$$

$$W = \frac{\hbar}{2\omega_e}(\omega^2 + \omega_e^2 + c^2\mathbf{k}^2)|u|^2 \qquad (33)$$

or taking into account (24)

$$W = \hbar\omega\frac{\omega}{\omega_e}|u|^2 \qquad (34)$$

$$\mathbf{P} = \hbar\mathbf{k}\frac{\omega}{\omega_e}|u|^2 \qquad (35)$$

For a planar electron wave (23) (31) - (35), the relations follow

$$W = -\frac{\hbar\omega}{e}\rho \qquad (36)$$

$$\mathbf{P} = \frac{\mathbf{k}}{\omega}W \qquad (37)$$

$$\mathbf{j} = -\frac{ec^2}{\hbar\omega}\mathbf{P} \tag{38}$$

Let us consider some arbitrary "portion" of the planar electron wave, located in an allocated volume $V$. I intentionally use the term "portion" instead of the term "quantum" to avoid undesirable associations with the "indivisible quanta" that are considered in quantum mechanics. According to the definition of energy-momentum tensor and the electric charge density, this "portion" of the wave has the energy

$$E = \int_V W d\mathbf{r} \tag{39}$$

the momentum

$$\mathbf{p} = \int_V \mathbf{P} d\mathbf{r} \tag{40}$$

and the electric charge

$$q = \int_V \rho d\mathbf{r} \tag{41}$$

In particular, for a planar wave (23), taking into account expressions (31), (34) and (35) one obtains

$$q = -eZ \tag{42}$$

$$E = \hbar\omega Z \tag{43}$$

$$\mathbf{p} = \hbar\mathbf{k}Z \tag{44}$$

where

$$Z = V\frac{\omega}{\omega_e}|u|^2 \tag{45}$$

Taking into account the transformation properties of the frequency $\omega$ and the volume $V$, it is easy to see that $Z$ is a 4-scalar.

The energy and the momentum of the same "portion" of the wave are connected by the relation

$$\mathbf{p} = \frac{E}{\hbar\omega}\hbar\mathbf{k} \tag{46}$$

If some portion of the electron wave with energy $E$ has been absorbed by some object, then the object simultaneously obtained momentum $\mathbf{p}$, as determined by expression (46). Similarly, if an object emits a certain portion of the electron wave with energy $E$, it simultaneously loses momentum $\mathbf{p}$, as defined by expression (46). Thus, the absorbed or emitted portion of the electron wave will be perceived as a particle with energy (43), momentum (44) and electric charge (42).

Taking into account expressions (24), (43) and (44), one can write the relation between the energy and the momentum of any "portion" of the planar electron wave (23):

$$E^2 = M_0^2 c^4 + c^2 \mathbf{p}^2 \qquad (47)$$

where

$$M_0 = \frac{\hbar \omega_e}{c^2} Z \qquad (48)$$

Expression (47) is similar to the relativistic relation connecting the energy and the momentum of a classical particle. In this case, parameter $M_0$ plays the role of the rest mass of the "particle". More specifically, if the planar electron wave will be absorbed or emitted as the portions with identical energy

$$E = \hbar \omega \qquad (49)$$

and the momentum of these portions will be connected with the wave number of electron wave (23) by de Broglie ratio (1)

$$\mathbf{p} = \hbar \mathbf{k} \qquad (50)$$

For such "portions" $Z = 1$, each will have an electric charge

$$q = -e \qquad (51)$$

For example, a proton with electric charge $e$ can capture a "portion" of an electron wave with a total charge of $-e$ and thus can become a neutral hydrogen atom.

Such "portions" of the electron wave will be perceived in experiments because particles with a rest mass $m_e$ are determined by ratio (5). Precisely these portions of an electron wave—which are entirely emitted or conversely, absorbed by any object—can be interpreted as the particles-electrons, although our analysis contains no particles, considering only the continuous electron wave.

Based on this analysis and the results of [17-19], one can formulate the following statement, which stands in opposition to de Broglie hypothesis (let us call it conditionally "anti-de Broglie"): any continuous wave (electromagnetic, electron) involving an interaction with matter having an atomic structure can manifest itself as a flux of particles with momentum $\mathbf{p} = \hbar \mathbf{k}$, where $\mathbf{k}$ is the wave vector characterizing the wave. If the de Broglie hypothesis contradicted all of the classic representations about waves and particles and introduced the wave-particle duality into physics as a fundamental property of whole matter, the "anti-de Broglie" statement (as will be shown below both in this paper and in the following papers in this series) simply notes that the results of the continuous wave's interaction with the discrete atoms of the detector is similar to the interaction of the flux of the particles (quanta) with the detector, thus leading to misinterpretation of the discrete events. This statement does not lead to physical paradoxes and is consistent with the representations of classical physics. Such a perspective turns quantum mechanics into the usual theory of the classical field (electron and electromagnetic), in which there is nothing except for the fields, and all phenomena should be interpreted from this perspective.

In the long-wave (Schrödinger) approximation (25), the expressions (16), (17), (21) and (22) take the form

$$\rho = -e|\psi|^2 \tag{52}$$

$$\mathbf{j} = i\frac{ec^2}{2\omega_e}\left(\psi^*\nabla\psi - \psi\nabla\psi^* + i2\alpha_0\mathbf{A}\psi\psi^*\right) \tag{53}$$

$$W = \hbar\omega_e\psi\psi^* + i\frac{\hbar}{2}\left(\psi^*\frac{\partial\psi}{\partial t} - \psi\frac{\partial\psi^*}{\partial t}\right) + \hbar\alpha_0 c\varphi\psi\psi^* +$$
$$+ \frac{\hbar c^2}{2\omega_e}(\nabla\psi^* - i\alpha_0\mathbf{A}\psi^*)(\nabla\psi + i\alpha_0\mathbf{A}\psi) + \frac{1}{8\pi}(\mathbf{E}^2 + \mathbf{H}^2) \tag{54}$$

$$\mathbf{P} = i\frac{\hbar}{2}(\psi\nabla\psi^* - \psi^*\nabla\psi) + \hbar\alpha_0\mathbf{A}\psi\psi^* + \frac{1}{4\pi c}[\mathbf{E}\mathbf{H}] \tag{55}$$

where the wave function $\psi$ satisfies the Schrödinger equation.

Here, it is assumed that

$$\left|\frac{\partial\psi}{\partial t}\right| \ll \omega_e|\psi| \tag{56}$$

and

$$\alpha_0 c|\varphi| \ll \omega_e \tag{57}$$

Relation (56) expresses the condition of a slow change in amplitude $\psi$ of wave function (6), whereas from the corpuscular perspective, condition (57) is interpreted as the fact that a particle's potential energy in an external electric field is substantially less than its own energy (the rest energy). From a considered perspective, the latter interpretation is superfluous and the relation (57) is the only restriction imposed on the external electric field in which the Schrödinger approximation (25) works.

## 2.3 Maxwell-Pauli field. The spin

In general, the electron field is described by the Dirac equation that plays a role of the field equation. In the long-wave (Schrödinger) approximation (25), (56), (57), the Dirac equation reduces to the Pauli equation with respect to the two-component spinor wave function $\Psi = \begin{pmatrix} \psi_1 \\ \psi_2 \end{pmatrix}$. The Pauli equation is more appropriate for our purposes than the Dirac equation because allows a visual presentation of the property of the electron waves that we call the "spin" from the standpoint of classical physics.

The equations of the Maxwell-Pauli field can be obtained from the variational principle (7), where the Lagrangian density

$$\mathscr{L} = -\frac{\hbar}{2\omega_e}\left[(\nabla\Psi^\dagger - i\alpha_0\mathbf{A}\Psi^\dagger)(\nabla\Psi + i\alpha_0\mathbf{A}\Psi) + i\omega_e\left(\frac{\partial\Psi^\dagger}{\partial t}\Psi - \Psi^\dagger\frac{\partial\Psi}{\partial t}\right) \right. \\ \left. -2\omega_e c\alpha_0\varphi\Psi^\dagger\Psi + \alpha_0\Psi^\dagger\boldsymbol{\sigma}\mathbf{H}\Psi\right] + \frac{1}{8\pi}(\mathbf{E}^2 - \mathbf{H}^2) \quad (58)$$

$\alpha_0 = \dfrac{e}{\hbar c} > 0$ is the constant of interaction, $\boldsymbol{\sigma}$ are the Pauli matrices,

$$\mathbf{H} = \mathrm{rot}\mathbf{A}, \ \mathbf{E} = -\frac{1}{c}\frac{\partial\mathbf{A}}{\partial t} - \nabla\varphi \quad (59)$$

Variation of Eqs. (7) and (58) with respect to $\Psi^\dagger$ leads to the Pauli equation for electron

$$i\frac{1}{c}\frac{\partial\Psi}{\partial t} = \left[\frac{c}{2\omega_e}\left(\frac{1}{i}\nabla + \alpha_0\mathbf{A}\right)^2 - \alpha_0\varphi + \frac{\alpha_0 c}{2\omega_e}\boldsymbol{\sigma}\mathbf{H}\right]\Psi \quad (60)$$

or in a more familiar form,

$$i\hbar\frac{\partial\Psi}{\partial t} = \left[\frac{1}{2m_e}\left(\frac{\hbar}{i}\nabla + \frac{e}{c}\mathbf{A}\right)^2 - e\varphi + \frac{e\hbar}{2m_e c}\boldsymbol{\sigma}\mathbf{H}\right]\Psi \quad (61)$$

where the "mass of the electron" (5) is introduced for convenience only. Variation (7), (58) with respect to the vector and the scalar potentials of the electromagnetic field ($\varphi, \mathbf{A}$) leads to the Maxwell equations

$$\mathrm{rot}\,\mathbf{H} = \frac{1}{c}\frac{\partial\mathbf{E}}{\partial t} + \frac{4\pi}{c}\mathbf{j} \quad (62)$$

$$\mathrm{div}\,\mathbf{E} = 4\pi\rho \quad (63)$$

where

$$\rho = -e\Psi^\dagger\Psi \quad (64)$$

$$\mathbf{j} = -\frac{e\hbar}{2m_e i}\left(\Psi^\dagger\nabla\Psi - (\nabla\Psi^\dagger)\Psi\right) - \frac{e^2}{m_e c}\mathbf{A}\Psi^\dagger\Psi - \frac{e\hbar}{2m_e}\mathrm{rot}\,(\Psi^\dagger\boldsymbol{\sigma}\Psi) \quad (65)$$

The second couple of Maxwell equations are a consequence of definition (59):

$$\mathrm{rot}\,\mathbf{E} = -\frac{1}{c}\frac{\partial\mathbf{H}}{\partial t} \quad (66)$$

$$\text{div}\mathbf{H} = 0 \tag{67}$$

Equations (62), (63), (66) and (67) are the Maxwell equations of the classical electromagnetic field, therefore parameter $\rho$ remains the electric charge density of the electron field, whereas while vector **j** is the electric current density of the electron field. From here, it again follows that *the electron field has continuously distributed the electric charge and the electric current in space* that cannot be reduced to the motion of charged particles.

The variation with respect to $\Psi$ leads to the conjugate Pauli equation with respect to the wave function $\Psi^\dagger$ and does not provide anything new compared to equation (59). Thus, the system of equations (59)-(67) describes the nonlinear, self-consistent Maxwell-Pauli field by taking into account the interaction of the classical electron and electromagnetic fields. As shown in [23-25], Pauli equation (60), (61) can be written in hydrodynamic form:

$$\frac{\partial \rho}{\partial t} + \text{div}\mathbf{j} = 0 \tag{68}$$

$$m_e \frac{d\mathbf{v}}{dt} = \mathbf{K} + \nabla \Pi^* - \frac{e}{m_e c} S_k \nabla H_k^* \tag{69}$$

$$\frac{d\mathbf{S}}{dt} = -\frac{e}{m_e c} \mathbf{S} \times \mathbf{H}^* \tag{70}$$

where we have introduced the notations: $k = 1,2,3$ is the vector index,

$$\frac{d}{dt} = \frac{\partial}{\partial t} + (\mathbf{v}\nabla) \tag{71}$$

$$\mathbf{j} = \mathbf{j}_0 + c\,\text{rot}\,\mathfrak{m} \tag{72}$$

$$\mathbf{j}_0 = -\frac{e\hbar}{2m_e i}\left(\Psi^\dagger \nabla \Psi - (\nabla \Psi^\dagger)\Psi\right) - \frac{e^2}{m_e c}\mathbf{A}\Psi^\dagger \Psi \tag{73}$$

$$\mathbf{v} = \mathbf{j}_0/\rho \tag{74}$$

$$\mathbf{K} = -e\left(\mathbf{E} + \frac{1}{c}\mathbf{v} \times \mathbf{H}\right) \tag{75}$$

$$\mathfrak{m} = -\frac{e\hbar}{2m_e c}\Psi^\dagger \boldsymbol{\sigma} \Psi \tag{76}$$

$$\mathbf{S} = \frac{\hbar}{2}\frac{\Psi^\dagger \boldsymbol{\sigma} \Psi}{\Psi^\dagger \Psi} \tag{77}$$

$$\mathbf{H}^* = \mathbf{H} + \mathbf{H}' \tag{78}$$

$$H_k' = -c\nabla\left(\frac{1}{\rho}\nabla s_k\right) \tag{79}$$

$$\Pi^* = \frac{e^2}{2m_e\rho^2}|\nabla\mathbf{s}|^2 \tag{80}$$

$$|\nabla\mathbf{s}|^2 \equiv (\nabla s_k \nabla s_k) \tag{81}$$

$$\mathbf{s} = \frac{\hbar}{2}\Psi^\dagger\boldsymbol{\sigma}\Psi \tag{82}$$

In view of the properties of the rot operator, from (68), (72) and (74), one obtains

$$\frac{\partial\rho}{\partial t} + \mathrm{div}\rho\mathbf{v} = 0 \tag{83}$$

From a formal perspective, equations (68) - (70) describe the flow of an incompressible electrically charged magnetic fluid in an external electromagnetic field. Indeed, equation (69) is the Euler equation, which describes the fluid flow with velocity $\mathbf{v}$. This fluid has a mass density

$$\rho_m = m_e\Psi^\dagger\Psi, \tag{84}$$

electric charge density $\rho$ (64) and an internal magnetic moment continuously distributed in space with density $\mathfrak{m}$, which is not connected with a flow of the passage fluid. In this case, each element $dV$ of the charged magnetic fluid in an external electromagnetic field is subject to the action of the Lorentz force $\mathbf{K}\Psi^\dagger\Psi dV$ and the force $-\frac{e}{m_e c}S_k\nabla H_k^*\Psi^\dagger\Psi dV = \mathfrak{m}dV\nabla\cdot\mathbf{H}^*$, which is caused by the action of the effective magnetic field $\mathbf{H}^*$ on the magnetic moment $\mathfrak{m}dV$ of the considered element of the fluid. Both of these forces are absolutely classical, if the electric charge and the internal magnetic moment are considered classical.

From the relations (64) and (84), it follows that the ratio of the electric charge density to the mass density of the magnetic fluid is the same at all points of space

$$\frac{\rho}{\rho_m} = -\frac{e}{m_e} \tag{85}$$

and is equal to the specific charge of the point "electron". Taking this into account, the equation (68) is both the continuity equation for the magnetic fluid and the law of conservation of its electric charge.

Equation (70) can be written as

$$\Psi^\dagger\Psi\frac{d\mathbf{S}}{dt} = \mathfrak{m}\times\mathbf{H}^* \tag{86}$$

The value $\mathfrak{m}dV\times\mathbf{H}^*$ is equal to the torque acting on the magnetic moment $\mathfrak{m}dV$ of the element $dV$ of the magnetic fluid, so equation (70), (86) describes the change in the internal angular momentum of the magnetic fluid that is unconnected to its flow. From equation (86), by taking into account continuity equation (68), one arrives at the conclusion that such a magnetic fluid possesses an internal angular momentum continuously distributed in space with a density **s** (82), and this angular momentum is an intrinsic property of the fluid.

The magnetic fluid has energy that is continuously distributed in space with density [25]

$$W = \Psi^\dagger\Psi\left(\frac{1}{2}m_e\mathbf{v}^2 - \Pi^*\right) + \rho\varphi - \mathfrak{m}\mathbf{H}^* + \frac{\hbar^2}{4m_e}\Delta\rho \qquad (87)$$

where $\frac{1}{2}m_e\mathbf{v}^2$ describes the energy connected with the motion of the electron wave, $\rho\varphi$ describes the potential energy of the distributed charge of the electron wave in an electric field, and $-\mathfrak{m}\mathbf{H}^*$ describes the potential energy of the distributed internal magnetic moment of the electron wave in an external magnetic field. These terms are classical in nature. The remaining terms of equation (87) have no classical analogue and are specific only to the electron waves. Note that the final term in equation (87) for spatially localized electron waves (e.g., in hydrogen atom) can be discarded because it disappears in the integration over the space.

We see that the electron wave in an external magnetic field described by the Pauli equation behaves like a classically charged magnetic fluid (magnetic continuous medium), possessing the internal angular momentum and magnetic moment continuously distributed in space. This enables consideration of the electron wave as a classical continuous field that has

- an electric charge, continuously distributed in space with density $\rho$,

- internal angular momentum, continuously distributed in space with density **s** and not connected with the motion of the electron wave,

- an internal magnetic moment, continuously distributed in space with density $\mathfrak{m}$ and unconnected to the motion of the electric charges of the electron wave,

- an electric current, continuously distributed in space with density **j**, which consists of the current with density $\mathbf{j}_0$ caused by the motion of the electric charges of an electron wave and the current with density $c\,\mathrm{rot}\,\mathfrak{m}$ connected to the internal magnetic moment of the electron wave $\mathfrak{m}$,

- energy continuously distributed in space with density $W$, and

- momentum continuously distributed in space with density

$$\mathbf{P} = \Psi^\dagger\Psi m_e\mathbf{v} = \frac{\hbar}{2i}\left(\Psi^\dagger\nabla\Psi - (\nabla\Psi^\dagger)\Psi\right) + \frac{e}{c}\mathbf{A}\Psi^\dagger\Psi \qquad (88)$$

Note that the same conclusion can be made based on an analysis of the relativistic, hydrodynamic forms of the Dirac equation [26-28].

Thus, we see that the classical electron field described by the Dirac equation (or in a simplified form, by the Pauli equation) is fundamentally different from the classical electromagnetic field (which does not have an electric charge, an internal angular momentum and an internal magnetic moment continuously distributed in space).

From the properties of the Pauli matrices, it follows that vector **S** has a constant length

$$|\mathbf{S}|^2 = \frac{\hbar^2}{4} \qquad (89)$$

From expressions (76) and (82), one obtains

$$\mathfrak{m} = \gamma_e \mathbf{s} \qquad (90)$$

where

$$\gamma_e = -\frac{e}{m_e c} \qquad (91)$$

From the above analysis, it follows that any elementary volume $dV$ of the electron wave has electric charge

$$dq = \rho dV \qquad (92)$$

internal angular momentum

$$d\mathbf{L}_s = \mathbf{s} dV \qquad (93)$$

and internal magnetic moment

$$d\boldsymbol{\mu} = \mathfrak{m} dV \qquad (94)$$

Property (93) of the electron field corresponds to what we call the "spin". Taking into account expression (90), for any "portion" of the electron field, one obtains

$$d\boldsymbol{\mu} = \gamma_e d\mathbf{L}_s \qquad (95)$$

It follows from (95) that the internal gyromagnetic ratio anywhere in the electron field is the same and equal to $\gamma_e$. According to (91) the internal gyromagnetic ratio for the electron wave is 2 times more than the gyromagnetic ratio connected with the motion of the electric charges. This again indicates that the internal angular momentum and the magnetic moment are intrinsic properties of the electron field and cannot be reduced to the motion of electric charges.

Using expressions (92) and (93), one obtains

$$d\mathbf{L}_s = (-dq/e)\mathbf{S} \qquad (96)$$

From here, in view of (89), it follows that

$$|d\mathbf{L}_s| = \frac{\hbar}{2} \frac{|dq|}{e} \qquad (97)$$

In other words, the internal angular momentum (and thus, the magnetic moment) of the electron field contained in any elementary volume $dV$ is proportional to the electric charge of the electron wave contained in this volume.

More specifically, let us consider a "portion" of the electron wave with an electric charge equal to $-e$, for example, the electron wave in the hydrogen atom. According to (95) and (97), this portion will have internal angular momentum $\frac{1}{2}\hbar$ and internal magnetic moment $\mu = \mu_B$, where $\mu_B = \frac{e\hbar}{2m_e c}$ is the Bohr magneton. As shown above, the absorption or emission of such a "portion" by any object will behave like a classical particle with a rest mass $m_e$. Because it also has the own angular momentum $\frac{1}{2}\hbar$, it will obviously be "exchanged" by the angular momentum with the object that either absorbs or emits it, in full compliance with classical mechanics. Thus, a "portion" of the electron wave will have all of the properties of the particle: the electron. Of course, although such a portion can be called "an electron", we must remember that in reality, we consider only an imaginary (i.e., a mentally allocated) "portion" of the continuous classical electron wave.

## 3 The interaction of an electron wave with matter

### 3.1 "Wave-particle duality" in the interference of an electron wave

The interaction of the electron wave with an atom can be characterized by the differential cross-section of the process, which is defined as

$$d\sigma = \frac{dN}{j_p} \qquad (98)$$

where $dN$ is the number of events (for example, the excitations of the atoms, the clicks of a detector, the spots on a detecting screen) that occur in unit time under the action of the incident electron wave and $j_p$ is the parameter associated with the electric current density of the electron wave (17) by the relation

$$j = -ej_p \qquad (99)$$

If we interpret the electron wave not as a continuous classical field but as a flux of the particles (i.e., the electrons) with the same electric charge $-e$, then parameter $j_p$ will mean the electron flux density, i.e. the number of "particles" passing per time unit through a unit cross-section of the beam.

The total effective cross-section of the considered event is obtained by integrating (for continuous events) or summing (for discrete events) the expression (98) over all possible realizations event:

$$\sigma = \frac{N}{j_p} \quad (100)$$

where $N$ is the number of all possible considered events that took place per unit time. Let us consider as an example the excitation (ionization) of a detector's atoms by the electron wave, considering the latter as a classical wave similar to the classical electromagnetic wave. Further, the electron wave is considered in the long-wave (Schrödinger) approximation (25) (56) (57). We assume that the incident electron wave is either planar or "close to planar", similar to, e.g., an electron wave passing through a double slit. In this case, from expression (53) at $\mathbf{A} = 0$, one obtains

$$j_p = v|\psi|^2 \quad (101)$$

where $v$ is the group velocity of the incident electron wave and $\psi$ is the wave function that describes the wave and satisfies the Schrödinger equation. Here, the wave function is not normalized to unity, as is usual in quantum mechanics; instead, it determines the actual amplitude of the electron wave; in particular, value (52) calculated by using the wave function, is equal to the density of the electric charge in the incident electron wave. It is assumed that the electric charge of the electron wave is continuously distributed in space.

Using expressions (100) and (101), one can write

$$N = \sigma v |\psi|^2 \quad (102)$$

Cross-sections (98) and (100) are calculated by standard methods of quantum scattering theory [29], using the assumption $|\psi|^2 = 1$. For this reason, values $d\sigma$ and $\sigma$ are independent of $|\psi|^2$, although they depend on velocity $v$, and for any considered event they are a characteristic of that event.

Let us consider, for example, an "inelastic" interaction of the electron wave with an atom (I deliberately avoid the standard term "inelastic collision of electrons with an atom" used in quantum mechanics), at which the excitation (ionization) of the atom that was previously in the ground state occurs.

For example, the total cross-section of the excitation of the first excited state of the hydrogen atom ($n = 2$) is as follows [29]:

$$\sigma_2 = \frac{4\pi}{v^2} 0.555 \ln \frac{v^2}{0.50} \quad (103)$$

Note that at the calculation of $d\sigma$ using the methods of quantum mechanics, the concept of "a particle" is not used anywhere: it solves the wave equation (e.g. Schrödinger equation) describing a continuous (and, so the classical) field, after which $d\sigma$ and $\sigma$ are calculated by means of the found solution. Particles, e.g., electrons, in the quantum theory "appear" only at the stage of interpreting solutions to the wave equations from a corpuscular perspective. However, as will be shown further, there is no need for such an interpretation because the purely wave (in the classical sense) perspective on electron waves enables an explanation of the discrete nature of their interaction with matter (e.g., detector).

Accepting that excitation of the atoms of a detector under the action of the incident electron wave occurs randomly, one can write

$$N = N_0 w$$

where $w$ is the probability of excitation of a detector's arbitrary atom per unit time and $N_0$ is the total number of the detector's atoms. If the detector is considered spatially uniform, then $N_0$ will be the same at all points of the electron wave's impact on the detector and can be considered as a constant characteristic of the detector.

In this case, expression (102) can be rewritten in the form

$$w = b|\psi|^2 \tag{104}$$

where

$$b = \sigma v / N_0 \tag{105}$$

does not depend on the amplitude of incident electron wave $|\psi|^2$.

Expression (104), which determines the probability of the excitation of an atom by the electron wave per unit time is similar to Fermi's golden rule, which determines the probability of an atom's excitation by light (electromagnetic) waves. In (104), however, the role of the square of the modulus of the electric vector of the electromagnetic wave plays the square of the modulus of the wave function of the incident electron wave $|\psi|^2$.

We assume that the excitation of atoms by electron wave is manifested in the form of the detector clicks or appearance of spots on the detecting screen.

In this case, the probability that a randomly chosen atom being in the electron wave field will be excited during time $t$ is defined by the expression [17-19]

$$P_+ = 1 - \exp(-wt) \tag{106}$$

Taking into account (104), one obtains

$$P_+ = 1 - \exp(-b|\psi|^2 t) \tag{107}$$

where wave function $\psi$ is a solution of the wave equation, for example, the Schrodinger equation for the corresponding problem.

Using expression (107), it is easy to calculate the process of the excitation of the atoms on the detecting screen under the action of a continuous classical electron wave. By analogy with the process of the interaction of light waves and the detecting screen [17-19], this calculation can be performed using the Monte Carlo method. Consequently, we will see the random appearance of discrete spots on the detecting screen: i.e., the excited atoms. Such a calculation can be executed for a double-slit experiment with a very weak electron wave [1]; in this case, the calculated interference pattern build-up will be identical to that obtained in [17-19] for light. For this reason, I will not provide the figures showing the interference pattern build-up in double-slit experiments with an electron wave, referring the reader to [17,18], in which such figures are presented and analyzed in detail. Comparison of the results of these calculations with the experimental data [1] shows their complete identity; in particular, a visualization of the results of the calculation in dynamics (see the supplemental material in [19]) completely reproduces a picture observed in experiment [20]. Once again, we note that these results were obtained considering the interaction of the classical electron wave with the detecting screen, which consists of discrete atoms. Therefore, in our reasoning, we never used such physical objects as particles (electrons), for which there is simply no need.

## 3.2 Born rule

However, if we want to interpret the discrete events (clicks of a detector, spots on the detecting screen) not as an excitation of a detector's individual atoms under the influence of continuous classical electron waves, but as a result of the particles-electrons' hit on the detector, we will have to consider the probability of finding an electron at a given point in space. In this case, assuming that hitting the "electron" on the detector inevitably causes a discrete event, this probability will coincide with the above-calculated probability of excitation of the detector's atoms by the continuous classical electron wave.

Let the total number of atoms in a selected volume $V$ of the material (e.g., on a selected surface of the detecting screen) be equal to $N$ and uniformly distributed thereon. Let us choose a small volume $dV$, which contains $dN = \frac{N}{V}dV$ atoms. Then, $dN_+ = P_+ dN$ atoms will be excited in volume $dV$ by time $t$:

$$dN_+ = P_+ \frac{N}{V} dV \tag{108}$$

Because here we interpret the excitation of the atoms as the hit of the particles-electrons, the number of electrons that reach volume $dV$ during time $t$ will be determined by expression (108). Accordingly, the probability of an electron hit to the selected volume $dV$ is

$$p\,dV = \frac{P_+}{\int P_+ dV} dV \tag{109}$$

where the integral takes over the entire volume of the material (or the entire surface of the detecting screen).

Thus, the probability density of finding the electron at a given point in space

$$p = \frac{P_+}{\int P_+ dV} \sim P_+ \tag{110}$$

Using expression (107), one obtains

$$p(t) = \frac{1 - \exp(-b|\psi|^2 t)}{\int [1 - \exp(-b|\psi|^2 t)] dV} \tag{111}$$

For short-term exposure satisfying the condition

$$b|\psi|^2 t \ll 1 \tag{112}$$

one obtains approximately

$$p = \frac{|\psi|^2}{\int |\psi|^2 dV} \tag{113}$$

or

$$p \sim |\psi|^2$$

that is none other than the Born rule for electrons (2).

For long-term exposure that does not satisfy condition (112), expression (111) can be rewritten as

$$\frac{p(t)}{p_0(t)} = \frac{1 - \exp(-b|\psi|^2 t)}{1 - \exp(-b|\psi_0|^2 t)} \tag{114}$$

where subscript "0" refers to a characteristic point on the detecting screen (e.g., in its centre for the double-slit experiment).

The above analysis leads to the following fundamental conclusions. First, the Born rule is a trivial consequence of quantum mechanics, if one considers the atomic (discrete) structure of matter and expression (104), which follows, for example, from the solution to the Schrödinger equation of the problem of inelastic scattering electron waves considered as a classical field by atoms. Second, there is no need to interpret the excitation of an atom as a result of a particle-electron's hit on the detecting screen, which means that there is no need to interpret the probability of the excitation of atoms under the influence of the electron wave as a probability of finding the electron at a given point. Third, the Born rule (2) (taking into account the above) is not a fundamental law of nature but instead has an approximate character—it is valid only for relatively weak electron waves and a relatively small exposure time, which satisfy condition (112), whereas for long-term exposure, it should be replaced by more general rule (111), (114). Moreover, for small exposure times (112), probability $p$ does not depend on the exposure time (time of observation), whereas for long-term exposure, the observed pattern of distribution of the excited atoms in space will change over time. Fourth, the Born interpretation (2) of wave function even for short-term exposures (112) takes place only in those cases in which a parameter $b$ can be considered constant for all space; this is possible when the phase velocity $v$ of the electron wave is the same for the entire wave (quasi-planar wave, as, e.g. in the double-slit experiment). If velocity $v$ is significantly different at different points of space (e.g., within the atom), then the Born interpretation (2) ceases to work even in the approximation (112).

Note that the direct application of the Born rule (2) to the interpretation of solutions to the wave equation and the experimental data is the cause of more than a century of conflict in physics that is called the wave-particle duality. The approach suggested above does not lead to such a conflict.

It can be said that quantum mechanics' corpuscular interpretation of discrete events arising out of the action of the continuous electron wave on a detector actually transfers a discrete structure of matter to the discrete structure of the electron wave. Instead of considering the interaction of a continuous (classical) electron wave with a substance having a discrete (atomic) structure, a fictitious system is considered in which the flux of discrete particles-electrons interact with structureless detector material.

Similarly, other "quantum" properties of "electrons" arise.

3.3 Heisenberg uncertainty relations

The Heisenberg uncertainty principle, along with the Born rule and the complementarity principle, is the basis of the Copenhagen interpretation of quantum mechanics. It is considered a quantitative justification of wave-particle duality achieved by interpreting uncertainty relations as natural constraints on the precision of simultaneous measurements of the position and momentum of a quantum particle.

The approach developed in this paper allows us to give a simple, classical explanation of Heisenberg's uncertainty principle by analogy with how this explanation was made for light [17,18].

Let us consider the classical wave packet (pulse) of the electron wave $\psi(\mathbf{r},t)$ with a finite dimension in space and time. This wave packet is considered a classical wave and represents a superposition of monochromatic waves with the wave numbers lying in a limited range. The distribution of the wave numbers in the wave packet is given by Fourier transform $\Phi(\mathbf{k},t)$ of the original wave field $\psi(\mathbf{r},t)$.

From the properties of the Fourier transform, which is the basis of an elementary proof of the uncertainty relation (see, e.g. [30,31]), it follows that

$$\Delta x \Delta k_x \geq \frac{1}{2} \qquad (115)$$

where $\Delta x$ is the characteristic width of the wave packet and $\Delta k_x$ is the characteristic width of the range of wave numbers $k_x$ of monochromatic waves entering into the packet. Similar relations can be written for the $y$ and $z$ coordinates. This is the "uncertainty relation" for classical waves. Obviously, with respect to the classical wave packet, the uncertainty relation (115) does not have the mystical meaning that is ascribed to the Heisenberg uncertainty principle in the Copenhagen interpretation of quantum mechanics because it does not impose restrictions on the measurement accuracy of the parameters of the quantum particle (which simply does not exist in our reasoning). Instead, it simply states a mathematical fact: the greater the width of the wave packet, the smaller the range of the wave numbers of monochromatic waves entering into the wave packet, and vice versa.

Multiplying expression (115) by the Planck constant and using expression (50), which is valid for the portion of the planar electron wave that has an electric charge equal to $-e$, we obtain the Heisenberg uncertainty relation (3):

$$\Delta x \Delta p_x \geq \frac{1}{2}\hbar \qquad (116)$$

and the similar relations for the $y$ and $z$ coordinates. Taking into account expression (50), these relations can be interpreted as a formal connection between the spatial width of the wave packet and a range of momentum of the "electrons" entering into the packet. If the energy of the classical wave packet $\psi(\mathbf{r},t)$ is less than the energy of one "electron" $\hbar\omega_e$, it is necessary to resort to the probabilistic interpretation (2) and expression (116) should be interpreted as the limitation on the accuracy of the simultaneous measurement of the position and momentum of the "electron". However, as we have seen, expression (116) was obtained formally from expression (115) for the classical wave field using expression (50), which was also obtained based on purely wave representations. Therefore, in reality expression (116) contains no more meaning than expression (115) and in fact is only another form of writing expression (115).

Similarly, it follows from the properties of Fourier transforms that if a wave packet has duration $\Delta t$, it contains a monochromatic wave of a certain frequency range, the width $\Delta \omega$ of which satisfies the relation

$$\Delta t \Delta \omega \geq \frac{1}{2} \qquad (117)$$

This expression is also devoid of any mystical meaning and means only one thing: the shorter the pulse of the electron wave, the wider the range of frequencies of the monochromatic electron waves that form this pulse. Using expression (49), inequality (117) can be formally written as the Heisenberg uncertainty relation for time and energy (4): $\Delta t \Delta E \geq \frac{1}{2} \hbar$, however, and this expression contains no more meaning than classical expression (117).

Thus, we conclude that from a considered perspective, the Heisenberg uncertainty principle reflects only the well-known properties (115) and (117) of the classical wave packets and in any case, it cannot be considered as the natural limitation on the measurement accuracy of the parameters of particles-electrons, which are simply absent from our reasoning.

## 4 The Compton effect

In quantum mechanics, the Compton effect occupies a special place. This effect is considered direct proof of the existence of photons. Exactly after the discovery of the Compton effect and its explanation based on photonic representations, many physicists began to perceive photons as real physical objects. The canonical (in quantum mechanics) explanation of the Compton effect considers an elastic scattering of photons by free electrons. In this case, photons are considered as massless, relativistic particles that have energy $\hbar \omega$ and momentum $\hbar \omega / c$, whereas the electron is considered as the point particles having the rest mass $m_e$.

However, even at the dawn of quantum mechanics, the Compton effect was explained without using the concept of the photon and electrons [32-38]: light was considered a classical electromagnetic wave, whereas electrons were described by the Klein-Gordon or Dirac equations and were also considered classical waves. Among the considered perspectives, the approach [34] based entirely on classical electrodynamics is of particular interest. In that paper, the classical electric current that creates the scattered electromagnetic wave was calculated based on the solution to the Klein-Gordon equation. Simultaneously, in papers [32,33], the wave equation was used only as a tool for calculating the matrix elements; this approach makes the obtained results more formal and less clear from a physical perspective. A pictorial explanation of the Compton effect was proposed by E. Schrödinger [37] based on purely wave representations: he considered light as a classical electromagnetic wave and drew an analogy between the scattering of this wave on the de Broglie wave and the Bragg scattering of light on ultrasonic waves, an issue that was considered by L. Brillouin.

Because I defend the view that there are neither photons nor electrons as particles, it is interesting to analyze the calculation of the Compton effect, which is considered in [34]. In this case, both waves (electromagnetic and electron) will be considered as a continuous classical wave field.

Full consideration of the Compton effect is possible only based on the relativistic Dirac equation [35,36]. Nevertheless, to understand why the mechanistic model (which considers an elastic collision of "photons" with "electrons") allows us to obtain the correct expression for Compton shift, the more simple Klein-Gordon equation will suffice.

According to classical electrodynamics, the scattering of an electromagnetic wave by a system of electric charges (discrete or continuously distributed in space) is a secondary radiation that is generated by these charges under the action of an incident electromagnetic wave. According to classical electrodynamics, the vector potential of electromagnetic radiation created by a system of electric charges at a long distance from the source is given by expression [21]

$$\mathbf{A}_s = \frac{1}{cR_0} \int \mathbf{j}(t - R_0/c + (\mathbf{rn})/c) dV \tag{118}$$

where $\mathbf{j}$ is the electric current density that creates the scattered radiation; $R_0$ is the distance from the centre of the electric charges to the observation point; $\mathbf{n}$ is the unit vector indicating the direction from the centre of the electric charges to the observation point; and $\mathbf{r}$ are the coordinates measured from the centre of the system of electric charges. In the case of Compton scattering, the source of secondary (scattered) radiation is the electric charge of the electron wave continuously distributed in space. Therefore, according to (17),

$$\mathbf{j} = i\frac{ec^2}{2\omega_e}\left(\Psi^*\nabla\Psi - \Psi\nabla\Psi^*\right) - \frac{e^2 c}{\hbar\omega_e}\mathbf{A}\Psi\Psi^* \tag{119}$$

Introduce

$$\mathbf{j} = \mathbf{j}_1 + \mathbf{j}_2 \tag{120}$$

where

$$\mathbf{j}_1 = i\frac{ec^2}{2\omega_e}\left(\Psi^*\nabla\Psi - \Psi\nabla\Psi^*\right) \tag{121}$$

$$\mathbf{j}_2 = -\frac{e^2 c}{\hbar\omega_e}\mathbf{A}\Psi\Psi^* \tag{122}$$

In the future, we will only consider the radiation corresponding to current (122).

For the weak incident electromagnetic wave (which corresponds to the usual conditions in the experiments on the Compton effect), when calculating (122) it is sufficient to confine only the linear approximation in which

$$\mathbf{A} \approx \mathbf{A}_0 \tag{123}$$

where

$$\mathbf{A}_0 = \frac{1}{2}\left(\mathbf{a}\exp[-i(\omega_0 t - \mathbf{k}_0\mathbf{r})] + \mathbf{a}^*\exp[i(\omega_0 t - \mathbf{k}_0\mathbf{r})]\right) \tag{124}$$

is the vector potential of the incident electromagnetic wave; $\mathbf{a}$ is its amplitude (constant vector); and $\mathbf{k}_0$ and $\omega_0$ are the wave vector and the frequency of the incident electromagnetic waves, respectively; wherein

$$|\mathbf{k}_0|^2 = \frac{\omega_0^2}{c^2} \tag{125}$$

$$(\mathbf{k}_0\mathbf{a}) = 0 \tag{126}$$

Here as usual, we select the gauge for the incident electromagnetic wave at which its scalar potential is equal to zero [21].

We assume that the electron wave is free ($\mathbf{A} = 0$), but nonmonochromatic:

$$\Psi = \int u(\mathbf{p})\exp\{-i[E(\mathbf{p})t - \mathbf{p}\mathbf{r}]/\hbar\}d\mathbf{p} \tag{127}$$

where function $u(\mathbf{p})$ describes the "spectrum" of the electron wave and function $E(\mathbf{p})$ satisfies the dispersion relation

$$E^2 = c^2\mathbf{p}^2 + m_e^2 c^4 \tag{128}$$

following from the Klein-Gordon equation for a free electron field.

Such a situation corresponds, for example, either to weakly coupled "outer electrons" in atoms or to a case in which an electron wave has passed through some field that has made it nonmonochromatic, and after this the electron wave is irradiated by a weak electromagnetic wave, its action on the electron wave can be ignored.

In this section, we will use parameters $E$ and $\mathbf{p}$ instead of a more natural wave frequency and wave vector. This is done, on the one hand, to avoid confusion between the designations of the wave characteristics of electromagnetic and electron waves and on the other hand, to clearly (as is habitual for all notations) to show how a law of the elastic collisions of "photons" and "electron" appears, which is the basis for the traditional explanation of the Compton effect.

Substituting (127) for (122), one obtains

$$\mathbf{j}_2 = -\frac{e^2 c}{2\hbar\omega_e}\mathbf{a}\int u(\mathbf{p}_1)u^*(\mathbf{p})\exp\{-i[(E(\mathbf{p}_1)-E(\mathbf{p})+\hbar\omega_0)t-(\mathbf{p}_1-\mathbf{p}+\hbar\mathbf{k}_0)\mathbf{r}]/\hbar\}d\mathbf{p}d\mathbf{p}_1 - \\
-\frac{e^2 c}{2\hbar\omega_e}\mathbf{a}^*\int u(\mathbf{p}_1)u^*(\mathbf{p})\exp\{-i[(E(\mathbf{p}_1)-E(\mathbf{p})-\hbar\omega_0)t-(\mathbf{p}_1-\mathbf{p}-\hbar\mathbf{k}_0)\mathbf{r}]/\hbar\}d\mathbf{p}d\mathbf{p}_1)$$
(129)

Let us introduce

$$\hbar\omega(\mathbf{p}_1,\mathbf{p}) = E(\mathbf{p}_1) - E(\mathbf{p}) + \hbar\omega_0 \tag{130}$$

$$\hbar\mathbf{k}(\mathbf{p}_1,\mathbf{p}) = \mathbf{p}_1 - \mathbf{p} + \hbar\mathbf{k}_0 \tag{131}$$

Then, (129) can be written in the form

$$\mathbf{j}_2 = -\frac{e^2 c}{2\hbar\omega_e}\mathbf{a}\int u(\mathbf{p}_1)u^*(\mathbf{p})\exp\{-i[\omega(\mathbf{p}_1,\mathbf{p})t - \mathbf{k}(\mathbf{p}_1,\mathbf{p})\mathbf{r}]\}d\mathbf{p}d\mathbf{p}_1 - \\
-\frac{e^2 c}{2\hbar\omega_e}\mathbf{a}^*\int u(\mathbf{p}_1)u^*(\mathbf{p})\exp\{i[\omega(\mathbf{p},\mathbf{p}_1)t - \mathbf{k}(\mathbf{p},\mathbf{p}_1)\mathbf{r}]\}d\mathbf{p}d\mathbf{p}_1)$$
(132)

Substituting (132) into (118), one obtains

$$\mathbf{A}_{s2} = -\mathbf{a}\frac{e^2}{2\hbar\omega_e}\frac{1}{R_0}\int u(\mathbf{p}_1)u^*(\mathbf{p})e^{-i\omega(\mathbf{p}_1,\mathbf{p})(t-R_0/c)}\int e^{i(\mathbf{k}(\mathbf{p}_1,\mathbf{p})-\mathbf{n}\omega(\mathbf{p}_1,\mathbf{p})/c)\mathbf{r}}dVd\mathbf{p}d\mathbf{p}_1 - \\
-\mathbf{a}^*\frac{e^2}{2\hbar\omega_e}\frac{1}{R_0}\int u^*(\mathbf{p}_1)u(\mathbf{p})e^{i\omega(\mathbf{p}_1,\mathbf{p})(t-R_0/c)}\int e^{-i(\mathbf{k}(\mathbf{p}_1,\mathbf{p})-\mathbf{n}\omega(\mathbf{p}_1,\mathbf{p})/c)\mathbf{r}}dVd\mathbf{p}d\mathbf{p}_1$$
(133)

is the component of the scattered radiation caused by the current $\mathbf{j}_2$.

Taking into account that $\int e^{i\mathbf{K}\mathbf{r}}dV = (2\pi)^3\delta(\mathbf{K})$ for any vector $\mathbf{K}$, where $\delta(\mathbf{K})$ is the delta function $\int \delta(\mathbf{K})d\mathbf{K} = 1$, one obtains

$$\mathbf{A}_{s2} = -\mathbf{a}\frac{(2\pi)^3 e^2}{2\hbar\omega_e}\frac{1}{R_0}\int u(\mathbf{p}_1)u^*(\mathbf{p})e^{-i\omega(\mathbf{p}_1,\mathbf{p})(t-R_0/c)}\delta(\mathbf{k}(\mathbf{p}_1,\mathbf{p})-\mathbf{n}\omega(\mathbf{p}_1,\mathbf{p})/c)d\mathbf{p}d\mathbf{p}_1 - \\
-\mathbf{a}^*\frac{(2\pi)^3 e^2}{2\hbar\omega_e}\frac{1}{R_0}\int u^*(\mathbf{p}_1)u(\mathbf{p})e^{i\omega(\mathbf{p}_1,\mathbf{p})(t-R_0/c)}\delta(\mathbf{k}(\mathbf{p}_1,\mathbf{p})-\mathbf{n}\omega(\mathbf{p}_1,\mathbf{p})/c)d\mathbf{p}d\mathbf{p}_1$$
(134)

Let us use the well-known property of the delta function

$$\delta(\mathbf{K}(\mathbf{p}_1)) = \frac{1}{J(\mathbf{p}_0)}\delta(\mathbf{p}_1 - \mathbf{p}_0) \tag{135},$$

where $\mathbf{p}_0$ is the solution (it is assumed the only solution) of the system of equations

$$\mathbf{K}(\mathbf{p}_1) = 0 \tag{136},$$

while

$$J(\mathbf{p}_1) = \left|\frac{\partial \mathbf{K}}{\partial \mathbf{p}_1}\right| \qquad (137)$$

is the Jacobian of transform $\mathbf{K}(\mathbf{p}_1)$, for which we assume that $J(\mathbf{p}_1) \neq 0$.

Then, denoting

$$\mathbf{K}(\mathbf{p}_1) = \mathbf{k}(\mathbf{p}_1,\mathbf{p}) - \mathbf{n}\omega(\mathbf{p}_1,\mathbf{p})/c \qquad (138)$$

in our case one obtains

$$\mathbf{A}_{s2} = -\mathbf{a}\frac{(2\pi)^3 e^2}{2\hbar\omega_e}\frac{1}{R_0}\int \frac{1}{J(\mathbf{p}_0)}u(\mathbf{p}_0)u^*(\mathbf{p})\exp\{-i\omega(\mathbf{p}_0,\mathbf{p})(t-R_0/c)\}d\mathbf{p} - \\ -\mathbf{a}^*\frac{(2\pi)^3 e^2}{2\hbar\omega_e}\frac{1}{R_0}\int \frac{1}{J(\mathbf{p}_0)}u^*(\mathbf{p}_0)u(\mathbf{p})\exp\{i\omega(\mathbf{p}_0,\mathbf{p})(t-R_0/c)\}d\mathbf{p} \qquad (139),$$

where $\mathbf{p}_0$ satisfies the equation

$$\mathbf{k}(\mathbf{p}_0,\mathbf{p}) - \mathbf{n}\omega(\mathbf{p}_0,\mathbf{p})/c = 0 \qquad (140)$$

Taking into account (140), definitions (130) and (131) can be written in the form

$$E + \hbar\omega = E_0 + \hbar\omega_0 \qquad (141)$$

$$\mathbf{p} + \mathbf{n}\hbar\omega/c = \mathbf{p}_0 + \hbar\mathbf{k}_0 \qquad (142),$$

where

$$E_0 = E(\mathbf{p}_0) \qquad (143)$$

$$E = E(\mathbf{p}) \qquad (144)$$

Function $E(\mathbf{p})$ satisfies the dispersion relation (128).

Because for fixed $\mathbf{n}$ and $\mathbf{k}_0$, there is a unique connection between vectors $\mathbf{p}_0$ and $\mathbf{p}$, expression (139) can be rewritten as

$$\mathbf{A}_{s2} = -\mathbf{a}\frac{(2\pi)^3 e^2}{2\hbar\omega_e}\frac{1}{R_0}\int \frac{\Delta(\mathbf{p}_0)}{J(\mathbf{p}_0)}u(\mathbf{p}_0)u^*(\mathbf{p})\exp\{-i\omega(\mathbf{p}_0)(t-R_0/c)\}d\mathbf{p}_0 - \\ -\mathbf{a}^*\frac{(2\pi)^3 e^2}{2\hbar\omega_e}\frac{1}{R_0}\int \frac{\Delta(\mathbf{p}_0)}{J(\mathbf{p}_0)}u^*(\mathbf{p}_0)u(\mathbf{p})\exp\{i\omega(\mathbf{p}_0)(t-R_0/c)\}d\mathbf{p}_0 \qquad (145),$$

where $\omega(\mathbf{p}_0) \equiv \omega(\mathbf{p}_0, \mathbf{p}(\mathbf{p}_0))$;

$$\Delta(\mathbf{p}_0) = \left|\frac{\partial \mathbf{p}}{\partial \mathbf{p}_0}\right| \qquad (146)$$

is the Jacobian of transform $\mathbf{p}(\mathbf{p}_0)$.

According to expression (145), the scattered radiation corresponding to current (122) will have frequencies $\omega(\mathbf{p}_0)$, which are determined by resolution of the system of equations (125), (128), (141)-(144) and depend on the angle of observation $\vartheta$, which by definition is equal to the angle between vectors $\mathbf{k_0}$ and $\mathbf{n}$:

$$\omega(\mathbf{p}_0) = \frac{E_0\omega_0 - c^2(\mathbf{k}_0\mathbf{p}_0)}{E_0 - c(\mathbf{p}_0\mathbf{n}) + \hbar\omega_0(1-\cos\vartheta)} \tag{147}$$

In this case,

$$\mathbf{p} = \mathbf{p}_0 + \hbar\mathbf{k}_0 - \mathbf{n}\hbar\omega(\mathbf{p}_0)/c \tag{148}$$

In particular, at $\mathbf{p}_0 = 0$, expression (147) reduces to the following well-known expression for Compton effect:

$$\frac{1}{\omega} - \frac{1}{\omega_0} = \frac{\hbar}{m_e c^2}(1-\cos\vartheta) \tag{149}$$

Thus, the current density (122) and scattered radiation (145) that are created by them correspond to Stokes scattering, i.e., to the Compton effect.

Using expression (145), it is not difficult to calculate the cross-section of the Compton scattering. Here, I will not do this, not only to avoid increasing the length of this paper but also so that this calculation can be found, for example, in [38], which enables us to ensure that it was actually performed within the framework of the classical field theory, whereas the particles (both "photons" and "electrons") appear only at the stage of interpretation. Similarly, the calculation of the Compton scattering can be done using the Dirac equation [35,36,38].

Note that expressions (125) (128) (141) (142), which are the basis of the considered explanation of the Compton shift (149), were obtained within the purely wave representations of electromagnetic and electron waves without the use of such concepts as "photon" and "electron". The "success" of the corpuscular explanation of the Compton shift is connected only to the fact that the dispersion relations (125), (128), (141) and (142) for interacting waves (electromagnetic and electron) are formally similar to the laws of conservation of energy and momentum for the elastic collision of two classical particles—i.e., "photon" and "electron".

From the above analysis, it becomes clear why classical electrodynamics, which considers electrons as point electric charges, cannot describe the Compton effect: the only component (122) of the current is responsible for Compton scattering; this component has no analogue for point electric charges considered in classical electrodynamics.

The above analysis shows that the Compton scattering should be considered as a purely classical effect, which the classical field theory is more than adequate to explain.

## 5 Concluding remarks

This paper shows that we have every reason to abandon electrons as real particles and instead consider the real classical physical field: an electron wave. Electron waves have unusual properties compared to electromagnetic waves: they have an electric charge, an intrinsic angular momentum and an intrinsic magnetic moment that are continuously distributed in space. This is unusual because the electric charge, the angular momentum and the magnetic moment are traditionally considered particle attributes because they emerged in classical physics' analysis of discrete objects.

Note that the angular momentum and magnetic moment distributed in space are the intrinsic properties of the electron wave, similar to its energy and momentum, and they cannot be reduced to any mechanical motion. That is why the spin gyromagnetic ratio of the "electron" (but, in fact, of any part of the electron wave) differs from the gyromagnetic ratio of the classical system of charged particles, which caused surprise among the founders of quantum mechanics and the desire to give this fact a mechanistic explanation, which (as we know) ended in failure.

Here, as a historical analogy, one can consider the known attempts to explain the laws of electrodynamics (in particular, to derive the Maxwell equations) based on mechanistic representations that were made in the late 19th century. Many famous physicists, especially A. Sommerfeld, worked in this direction. A. Sommerfeld's first work devoted to the electromagnetic theory appeared in 1892. In that work, Sommerfeld attempted to give a mechanical interpretation of the Maxwell equations based on a modified gyroscopic model of ether, as suggested by Kelvin. Although this paper caught the attention of Boltzmann, clear success was not achieved and thereafter, Sommerfeld adhered to the axiomatic approach to the construction of the fundamental equations of electrodynamics.

As a result of the development of classical field theory, an understanding developed that the electromagnetic field is a special kind of continuously distributed matter that has energy and momentum continuously distributed in space. This stopped further attempts to provide a mechanistic explanation of the electromagnetic field, and now we remember such explanations only as historical fact, curiosities in the development of physics. We are no longer surprised that there are non-mechanical forms of energy and momentum, nor that the energy and momentum of the classical electromagnetic field is continuously distributed in space and cannot be reduced to the motion of any particles. Similarly, we need to become accustomed to the fact that the electron field, along with energy and momentum, has an electrical charge, an intrinsic angular momentum and an intrinsic magnetic moment that are continuously distributed in space and cannot be reduced to the existence and motion of any particles.

Such a perspective turns quantum mechanics into the usual classical field theory in which there is nothing except fields, and all phenomena should be interpreted with these positions. From this perspective, quantum mechanics must be considered not as a development of mechanics, but as a purely field theory in the spirit of Maxwell.

Adopting this view, we will not be confronted by the contradictions that are characteristic of the wave-particle interpretation of quantum mechanics. Moreover, in this case the paradox associated with the infinity of electromagnetic energy of "electrons" disappears.

Note that from a formal perspective, the proposed theory is similar to that mentioned in the Introduction: the "Shut-up-and-calculate!" interpretation of quantum mechanics. However, if this interpretation urges us simply refuse any interpretation and "silently" solve the equations, in the proposed theory we introduce a new physical object—the electron field—that is considered to be a real, special form of matter, with quite certain physical properties.

The expected question is the following: How can we explain the fact that when considering the "electron" as a classic point particle having mass $m_e$ and electric charge $-e$, it is possible to explain the huge number of classical phenomena in plasma, in solids, etc.? First, we must remember that in classical systems, unlike in "quantum" systems, we never address single "electrons" because the properties of such systems are formed by macroscopic ensembles of "electrons". As is known, in the classical limitation, the Schrödinger equation becomes the Hamilton-Jacobi equation and the continuity equation. From a formal perspective, these equations collectively describe a classical ensemble of point particles with mass $m_e$ and electric charge $-e$, which can be called the Hamilton-Jacobi ensemble. This ensemble determines the macroscopic properties of classical systems. We can calculate the motion of such an ensemble in two ways: (i) by directly solving the Hamilton-Jacobi equation and the continuity equation, which is equivalent to the solution of the Schrödinger equation and the transition in the obtained solution to the classical limit; or (ii) by considering that the Hamilton-Jacobi equations have characteristics in the form of Hamilton equations that describe the motion of a single classical point particle with mass $m_e$ and electric charge $-e$ and therefore, solving the Hamilton equations for a set of classical particles of such an ensemble, one can find the motion of the ensemble as a whole. The latter circumstance explains the success of the classical theory of electrons in the description of phenomena in plasmas and solids when the "quantum" effects (the interference of electron waves) can be neglected.

Of course, although the approach suggested in this paper can seem radical, it is not new (see, e.g. [39]). It is fundamentally different from the traditional, habitual corpuscular perspective. However, as will be shown in the following papers in this series, this view allows us to describe the many so-called "quantum" effects—from uniform positions of classical field theory—in a natural way. Therefore, it will be shown that all of the basic properties of the atom, the photoelectric effect, thermal radiation and other "quantum" phenomena have a natural classical explanation without any quantisation.

# References

**Part 1**

**Part 2**